\documentclass[reprint,showpacs,amsfonts,amsmath,amssymb,prb]{revtex4-1}
\usepackage{graphicx}
\usepackage{dcolumn}
\usepackage{bm}

\begin{document}

\title{SU($N$) spin-wave theory: Application to spin-orbital Mott insulators}

\author{Zhao-Yang Dong}
\author{Wei Wang}
\author{Jian-Xin Li}
\email[]{jxli@nju.edu.cn}
\affiliation{National Laboratory of Solid State Microstructures and Department of Physics, Nanjing University, Nanjing 210093, China}
\affiliation{Collaborative Innovation Center of Advanced Microstructures, Nanjing University, Nanjing 210093, China}
\date{\today}

\begin{abstract}
We present the application of the SU($N$) ($N>2$) spin-wave theory to spin-orbital Mott insulators whose ground states exhibit magnetic orders. When taking both the spin and orbital degrees of freedom into account rather than projecting onto the Kramers doublet, the lowest spin-orbital locking energy levels, due to the inevitable spin-orbital multipole exchange interactions, the SU($N$) spin-wave theory should take the place of the SU($2$) one.
%The entanglement of the spin and orbital spaces due to the spin orbital coupling will lead to a rich variety of low energy effective Hamiltonians. The conventional treatment is to project onto the Kramers doublet, which is the lowest energy spin-orbital locking levels of the spin orbital coupling. However multipole-multipole exchange terms are inevitable if we take both degrees of freedom of spin and orbital into account.
To implement the application, we introduce an efficient general local mean field approach which involves all the local fluctuations into the SU($N$) linear spin-wave theory. Our approach is tested firstly by calculating the multipolar spin-wave spectra of the SU($4$) antiferromagnetic model.
Then we apply it to spin-orbital Mott insulators. It is revealed that the Hund's coupling would influence the effectiveness of the isospin-$1/2$ representation when the spin orbital coupling is not large enough. Besides, we also calculate the spin-wave spectra based on the first principle calculations for two concrete materials, $\alpha$-RuCl$_3$ and Sr$_2$IrO$_4$. The SU($N$) spin-wave theory appropriately depicts the low-energy magnons and the spin-orbital excitations qualitatively.
\end{abstract}

%\pacs{}

\maketitle

\section{INTRODUCTION}

The physics of transition-metal oxides (TMOs) with $4d$ or $5d$ orbitals occupied has drawn considerable attention recently.
One reason is that the spin-orbital coupling (SOC), which was considered as a small perturbation until recently, entangles the spin and orbital degrees of freedom.
This effect in cooperation with electronic correlations could give rise to a novel type of insulators (spin-orbital Mott insulators) in which the local moments are spin-orbital entangled $J_{\rm eff}=1/2$ Kramers doublets\cite{Witczak-Krempa2014,Kim2008,Kim2014}.
%Strong SOC gives rise to half-filled $J_{eff}= 1/2$ Kramers doublets together with the crystal field splitting of partially filled $d$ orbitals of transition ions, which forms the so-called spin-orbital Mott insulator with an effective spin-1/2 degree of freedom when a relatively small electronic correlation is presented\cite{Witczak-Krempa2014,Kim2008,Kim2014}.
Another is their crystal structures with a special bond geometry formed by edge-shared octahedra, which will result in the anisotropy and the frustration of the effective Hamiltonian\cite{Jackeli2009}, because the exchange coupling between the local moments depends highly on the spatial direction of the exchange path. The Hamiltonian with such a novel symmetry could lead to unconventional magnetism, including spin liquids, multipolar orders and uncommon magnetic orders\cite{Witczak-Krempa2014}. In real materials, the zigzag (Na$_2$IrO$_3$\cite{Liu2011} and $4d$ TMOs $\alpha$-RuCl$_3$\cite{Sears2015,Banerjee2016,Ran2017}), spiral (Li$_2$IrO$_3$ \cite{Biffin2014,Williams2016,Biffin2014a}) type magnetic orderings, and a canted antiferromagnetic (AF) structure (Sr$_2$IrO$_4$ )\cite{Gum2009,Kim2012} have been proved.

Generally, $4d$ and $5d$ states are spatially so extended that the Hubbard interaction is reduced compared to that of $3d$ states. However, owing to the large crystal field and SOC, a separate band with a reduced  bandwidth allows for the opening of a Mott gap. The underlying picture for this process is as following.
For a $d^{5}$ electronic configuration, when the two $e_g$ orbitals split off due to the crystal field of octahedrons, the five electrons loaded on the $t_{2g}$ orbitals results in a $s=1/2$ hole residing in an effective $l=1$ orbitals. A strong SOC leads to a system with a fully filled $J_{\rm eff}=3/2$ band and a half-filled $J_{\rm eff}=1/2$ band.
%The crystal field of the octahedron raises the $e_g$ orbitals and breaks the rotation symmetry, so the angular momentum $L$ is no longer the familiar conserved quantity in the atomic limit, but its projection onto the $t_{2g}$ orbitals. It behaves like a mirror angular momentum (whose operators obey $[L^\alpha,L^\beta]=-i\varepsilon^{\alpha\beta\gamma} L^\gamma$, where $\varepsilon^{\alpha\beta\gamma}$ is the Levi-Civita symbol) with quantum number $l=1$. Accordingly, an effective angular momentum $J_{\rm eff}=S-L$ is  defined in the $t_{2g}$ orbitals, in contrast to the total angular momentum $J=S+L$.
%Taking a large crystal field and SOC into account, five electrons on $d$ orbitals lead to a picture of a hole residing on the $J_{\rm eff}=1/2$ Kramers doublet.
Thus, the so-called spin-orbital Mott insulators emerge even with a relatively small electronic correlation. In this case, the $J_{\rm eff}=1/2$ states present the essential physics and effectively behavior as spin-$1/2$ pseudo spins. The resulting spin-exchange model can be obtained by projecting the electronic Hamiltonian onto the $J_{\rm eff}=1/2$ Kramers doublet which consists of only dipole-dipole interaction terms. To study the low-energy excitations of this spin-$1/2$ system with a magnetically ordered ground state, one can resort to the famous SU($2$) linear spin-wave theory\cite{Haraldsen2009}.
%And when the system is some magnetic ordered, it is simple and adequate to analysis the ground state by a local mean field theory. After minimizing the energy of the Hamiltonian, then we can turn to spin wave theory to study the low energy semiclassical dynamics of elementary excitations.
%Surely if the crystal field, SOC and Hubbard interation are strong enough, we can project the Hamiltonian onto the $J_{\rm eff}=1/2$ Kramers doublet, then it consists of only dipole-dipole interaction terms and the famous SU($2$) linear spin wave theory\cite{Haraldsen2009} is sufficient to deal with it.
%However, besides the collaboration, the competition between the crystal field and SOC will mix the $e_g$ and $t_{2g}$ orbitals, deviation from the spherical symmetry of the electron wave function brings the composition of $J_{\rm eff}=3/2$ into the Kramers doublet\cite{PhysRevB.91.241110}, and Hund's coupling in the multi-orbital system will induce electrons to orbit in the same direction. All of these would weaken the validity of the picture of a half-filling $J_{\rm eff}=1/2$ Kramers doublet.
However, in many real materials the mixing between the $e_g$ and $t_{2g}$ orbitals are always presented and the deviation from the spherical symmetry drags some composition of $J_{\rm eff}=3/2$ states into the Kramers doublet\cite{PhysRevB.91.241110}. In addition, the Hund's coupling in the multi-orbital system will induce electrons to orbit in the same direction. All of these would weaken the validity of the picture of a half-filling $J_{\rm eff}=1/2$ Kramers doublet, and complicate the spin exchange Hamiltonian by introducing the interactions between spin-orbital multipolar momentum\cite{Witczak-Krempa2014}.
Thus, the spin-orbital multipolar orders and excitations are needed to be considered.

Generally, to study a spin-$1/2$ system with a magnetically ordered ground state and small quantum fluctuations, the famous SU($2$) linear spin-wave theory\cite{Haraldsen2009} are used, in which the spins are regarded as a classical three-components vector and its fluctuations are described by rotations of the vector. However, when the degrees of freedom of both spins and orbitals are involved, it is insufficient to treat the local states as the rotations of a classical three-components angular momentum. Therefore, a generalization of the SU($2$) linear spin-wave theory is needed\cite{PhysRevB.60.6584}. Recently, the SU($N$) spin-wave theory based on the multi-boson approach has been introduced\cite{PhysRevB.85.125116,PhysRevLett.108.257203,PhysRevB.86.174428,Muniz2014}.
Since the generators of the SU($N$) group can be represented as bilinear forms in $N$-flavored bosons, instead of two bosons in the SU($2$) spin-wave theory, the low-energy modes of the SU($N$) spin-wave theory are described with $N-1$ different bosons, which would provide a more accurate description of the low-energy excitations for unconventional magnetic orders.

In this paper, we will use the SU($N$) spin-wave theory to study the magnetic excitations in spin-orbital Mott insulators. In the SU($N$) spin-wave theory, the local order parameter is defined in the space of SU($N$) unitary transformations of the local spin states, instead of the SU($2$) space of local spin rotations, and it consists of $N^2-1$ components of the SU($N$) order parameter in the most general form. Therefore, a universal local mean field theory facilitating the SU($N$) spin-wave theory is required. Here, we introduce a general efficient local mean field theory based on the supercoherent state\cite{Fatyga1991}, which fully includes the on-site quantum fluctuations essential for multipolar states.
As an illustration, we first apply the SU($N$) spin-wave theory to a toy three-band Hubbard model on a hexagon lattice, and focus on the examination of the effect of Hund's coupling by calculating the weights of $J_{\rm eff}=1/2$ stats in the ground state and spin-wave spectra.
%consisting of the excitations of the spin flipping within and jumping over the spin-orbital gap.
If the SOC is not large enough to lift the spin-orbital excitations across the $J_{\rm eff}=1/2$ and $J_{\rm eff}=3/2$ states away from those within the $J_{\rm eff}=1/2$ doublets, the Hund's coupling will compel the angular momentum $L$ to parallel the spin momentum. Therefore, the low energy physics is not governed only by the $J_{\rm eff}=1/2$ effective Hamiltonian. We then study the spin excitations in two systems of TMOs, $\alpha$-RuCl$_3$ and Sr$_2$IrO$_4$ where the effective Hamiltonian include both spin and orbital degrees of freedom, by using the SU ($N$) linear spin-wave theory. Our results for the magnetic ground states and their low-energy spin dynamics in two systems are consistent with recent experiments\cite{Banerjee2016,Ran2017,Kim2012,Kim2014}. In addition, we can obtain the high-energy spin-orbital excitations across the gap in the presence of the spin-orbital coupling. %Those in Sr$_2$IrO$_4$ are compared to the resonant inelastic X-ray scattering (RIXS) experiments, while those in $\alpha$-RuCl$_3$ wait for a comparison with future experiments.

The paper is organized in the following manner. In section \ref{spinwave}, we briefly review the Schwinger bosons representation and SU($N$) spin-wave theory, then introduce the general local mean field theory. In section \ref{su4}, based on the SU($4$) antiferromagnetic Hamiltonian\cite{Qi2008a,Wu2003,Hung2011}, we calculate its magnon excitations and spin-$3/2$'s $l=2$ multipole-multipole correlation function. In section \ref{tmo}, we apply the SU($N$) spin-wave theory to spin-orbital Mott insulators. First, we derive an effective Hamiltonian from a three-band Hubbard model with the SOC in the hexagon lattice and study the magnetic dynamics by the SU($N$) spin-wave theory.
Then we calculate the spin correlation function of $\alpha$-RuCl$_3$ with the five-band Hubbard model and correlation function of resonant inelastic X-Ray scattering (RIXS) operators\cite{Luo1993} of Sr$_2$IrO$_4$ with a three-band Hubbard model.

\section{SU($N$) linear spin-wave theory}\label{spinwave}

Muniz et al present a mathematical framework of the multi-boson approach to generalize the traditional spin-wave theory from SU($2$) to SU($N$)\cite{Muniz2014}. %The ferromagnetic and antiferromagnetic order of bilinear-biquadratic model have been analyzed in their paper.
As we know, the effective exchange models from the electron models in the strong interaction limit would always be written as %by SU($N$) generators, which can be represented as bilinear forms in $n$-flavored bosons,
\begin{equation}
  H_0=  J_{\mu\nu\mu'\nu'}^{rr'}S_r^{\mu\nu}S_{r'}^{\mu'\nu'}+h_{\mu\nu}^rS_r^{\mu\nu},
  \label{eq0}
\end{equation}
where the repeated index $r,r',\mu,\nu,\mu'\nu'$ is summed up, and $S_r^{\mu\nu}$ are the generators of SU($N$) group, which obey the commutation relations
\begin{equation}
[S_r^{\mu\nu},S_{r'}^{\mu'\nu'}]=\delta_{r,r'}(S_r^{\mu\nu'}\delta_{\mu'\nu}-S_r^{\mu'\nu}\delta_{\mu\nu'}).
\label{eq1}
\end{equation}
Then, they can be represented by Schwinger bosons. In the spin-wave theory, one of the bosons will be condensed depending on a given magnetic order and the rest $N-1$ different bosons will be used to describe the low-energy modes of systems.
In this section, we will first review the multi-boson approach based on the Schwinger bosons representation. Then, a general local mean field theory will be introduced and applied to the SU($N$) linear spin-wave theory.

\subsection{Schwinger bosons representation}

It is often useful to map a spin model into a bosonic one, which may be easier to study since bosons have simple commutation relations. Also, the common magnons are bosonic excitations which are proper to be represented in bosonic language. In the Schwinger bosons representation, the SU($N$) generators are written as\cite{Arovas1988},
\begin{eqnarray}
S_r^{\mu\nu}&=&b_r^{\mu\dagger} b_r^\nu,
\label{eq2}\\
\sum_{\mu=0}^{n-1} b_r^{\mu\dagger} b_r^\mu&=&n_b,
\label{eq3}
\end{eqnarray}
where $b_r^{\mu\dagger}$ and $b_r^\mu$ ($\mu=0,1,...,n-1$) are bosonic creation and annihilation operators on the local site $r$, respectively. Eq.~(\ref{eq3}) is a constraint on the bosonic operators in the physical space. $n_b$ is the number of bosons on the local site, denoting the order of the irreducible representations of SU($N$) group. For the well known SU($2$) linear spin-wave theory, we set $n_b=2S$. Here we use $n_b$=1 for simplicity. Thus, $n_b$ indicates the dimensions of the local state and there is an one-to-one match between each boson and each local dimension.
Furthermore, the space of local operators is a $n^2$-dimensional linear space, which could be expanded on the basis of the identity and the $n^2-1$ generators of SU($N$) group. Correspondingly, the identity is the constraint Eq.~(\ref{eq3}) and $n^2-1$ generators are bilinear forms $b^{\mu\dagger} b^\nu$. So, any local operator can be expressed as a linear combination of bosonic bilinear forms.

To sum up, all local fluctuations are described by bosonic particle-hole forms $b^{\mu\dagger} b^\nu$. For instance, if there is a local spin $S=3/2$, then local fluctuations can be expanded by the multipole expansion, which has $16=(2S+1)^{2}$ different scattering channels classified by the total spin of a pair of particle and hole.
\begin{equation}
  M_{l,m}=\sum_{m_1}(-1)^{s_2+m-m_1}C_{m_1,m-m_1,m}^{s_1,s_2,l} b^{s_1,m_1 \dagger} b^{s_2,m_1-m},
  \label{eq4}
\end{equation}
where $C_{m_1,m-m_1,m}^{s_1,s_2,l}$ are Clebsch-Gordan coefficients, and $(s_1,m_1), (s_2,m-m_1)$ are the spin quantum numbers of the particle and hole, respectively. $M_{l,m}$ is multipole spin operators. $M_{l,m}$ is the identity when $l=0$, the dipolar operators $S_+$, $S_-$ and $S_z$ when $l=1$, and the quadrupolar and octupolar operators when $l=2,3$. There are totally $16=\sum_{l=0}^3 2l+1$ multipole spin operators, which are equal to the dimensions of the space of local operators and can also be expanded by SU($N$) generators.
Therefore, SU($N$) spin-wave theory based on this multi-boson approach includes all of bosonic multipolar excitations.

\subsection{Local mean field theory}

It is necessary to construct a general local mean field theory to utilize all advantages of the SU($N$) spin-wave theory. As we known, the parameter manifold of a $n$-dimensional ($n$-D) state is $(n-1)$-D complex projective space CP($n-1$) when the overall phase is neglected. There are $n-1$ complex parameters, which are $2(n-1)$ real parameters. The local mean-field state should travel all over the space, so according to the supercoherent states constructed by Fatyga et al\cite{Fatyga1991}, we assume the test local wave function to be generated from a unitary transformation acting on an given state,
\begin{equation}
  \left|T\right\rangle_r=U(\bm{x}_r)b_r^{0\dagger}\left|0\right\rangle.
  \label{eq5}
\end{equation}
$U(\bm{x}_r)$ is the unitary transformation and $\left|0\right\rangle$ is the vacuum without any bosons:
\begin{eqnarray}
  U(\bm{x}_r) &=& \mathrm{exp}[i\sum_{\mu\neq0}(x_r^{2\mu-1}(b_r^{0\dagger} b_r^\mu+b_r^{\mu\dagger} b_r^0),
  \nonumber\\
  &&+x_r^{2\mu}(ib_r^{\mu\dagger} b_r^0-ib_r^{0\dagger} b_r^\mu))]
  \label{eq7} \\
  \left|0\right\rangle &=& (\underbrace{0,0,0,...,0}_n)^T,
\end{eqnarray}
where $\bm{x}\in \mathrm{R}^{2(n-1)}$, the $2(n-1)$-D real space. Obviously, $U(\bm{x}_r) $ is particle conserved, so the test state complies with the constraint Eq.~(\ref{eq3}). It is arduous to find the minimum in such a plain space. Thus, we will utilize the structure of CP($n-1$) to convert the $\bm{x}\in \mathrm{R}^{2(n-1)}$ parameter space to the rotation space in the $n$-D complex space,
\begin{align*}
  x^1&=\theta_1{\rm cos}(\theta_2){\rm cos}(\phi_1),  \\
  x^2&=\theta_1{\rm cos}(\theta_2){\rm sin}(\phi_1),  \\
  x^3&=\theta_1{\rm sin}(\theta_2){\rm cos}(\theta_3){\rm cos}(\phi_2),  \\
  x^4&=\theta_1{\rm sin}(\theta_2){\rm cos}(\theta_3){\rm sin}(\phi_2),  \\
  ...,\\
  x^{2n-3}&=\theta_1{\rm sin}(\theta_2)\ldots {\rm sin}(\theta_{n-1}){\rm cos}(\phi_{n-1}),  \\
  x^{2(n-1)}&=\theta_1{\rm sin}(\theta_2)\ldots {\rm sin}(\theta_{n-1}){\rm sin}(\phi_{n-1}),  \\
  \theta_j&\in\{0,\pi\},\phi_j\in\{0,2\pi\}.
\end{align*}
When $n=2$, it is the well known state of spin-$1/2$, $\left|T\right\rangle=({\rm cos}(\theta_1), e^{i\phi_1}{\rm sin}(\theta_1))^T$, where $(\theta_1,\phi_1)$ are Euler angles. It corresponds to a rotation in $2$-D complex space or $3$-D real space.

The mean field ground state of the system is the direct product state of local wave function, $\left|G\right\rangle=\bigotimes\left|T\right\rangle_r$, which would minimize the energy of $\left\langle G\right| H\left|G\right\rangle$. Due to the translational symmetry of the ground state, generally only the magnetic cell is considered in the spin-wave theory.

\subsection{SU($N$) Linear spin-wave approximation}

It is known that the spin-wave approximation is based on the Holstein-Primakoff (HP) bosons which define the spin-deviation operators. Its generalization can be obtained by extending the HP representation from SU($2$) to SU($N$)\cite{Muniz2014}. To obtain the SU($N$) HP bosons, we should first determine the condensed boson which creates the local state minimizing the mean-field energy. According to the variational form of the mean field ground state introduced in the last subsection, the condensed boson is the one minimizing $\left\langle G\right| H\left|G\right\rangle$, with $\left|G\right\rangle=\prod_r \tilde{b}_r^{0\dagger} \bigotimes\left|0\right\rangle_r$. It is related to the Schwinger boson $\mathbf{b}_r$ via the unitary transformation Eq.~(\ref{eq7}),
\begin{align}\label{eqb}
 \tilde{b}_r^{0\dagger}=\sum_\mu U_{0\mu}(\bm{x}_r)b_r^{\mu\dagger}.
\end{align}
Namely, $\tilde{b}_r^{0\dagger}$ is the $\mu=0$ component of $\tilde{\mathbf{b}}_r$, and the corresponding creation and annihilation operator are replaced by a number according to the constraint of Eq.~(\ref{eq3}),
%The SU($N$) HP bosons are obtained by condensing one of the Schwinger bosons straightforwardly according to the local mean field proximate ground state, $\left|G\right\rangle=\prod_r \tilde{b}_r^{0\dagger} \bigotimes\left|0\right\rangle_r$. $\tilde{b}_r^0$ are the bosons minimizing the ground state energy, $\tilde{\mathbf{b}}_r=U_r \mathbf{b}_r$, whose $\mu = 0$ boson is the one to be condensed: the corresponding creation and annihilation operator are replaced by a number according to the constraint of Eq.~(\ref{eq3}),
\begin{equation}
  \tilde{b}_r^{0^\dagger}\simeq\tilde{b}_r^0\simeq\sqrt{1-\sum_{\mu=1}^{n-1}\tilde{b}_r^{\mu\dagger} \tilde{b}_r^\mu}.
  \label{eq8}
\end{equation}
Then, the $N-1$ bosons $\tilde{b}_r^{\mu\neq0}$ become the HP bosons, which describe the spin waves originating from fluctuations around the ordered spin state created by the condensed boson $\tilde{b}_r^{0\dagger}$.
Substituting Eq.~(\ref{eq8}) into the Hamiltonian Eq.~(\ref{eq0}) and retaining only the quadratic terms, we get,
\begin{eqnarray}
  H & \simeq& \sum_{\langle r,r'\rangle}J_{0000}^{rr'}+(J_{\mu00\nu'}^{rr'}b_r^{\mu\dagger} b_{r'}^{\nu'}+J_{0\nu0\nu'}^{r,r'}b_r^\nu b_{r'}^{\nu'}+H.c)\nonumber\\
    &&+\sum_r  h_{00}^{r}+h_{\mu'\nu'}^{r} b_{r}^{\mu'\dagger} b_{r}^{\nu'}+\sum_{\langle r,r'\rangle}[(J_{\mu\nu00}^{rr'}-J_{0000}^{rr'} \delta_{\mu\nu})b_{r}^{\mu'\dagger} b_{r}^{\nu'}\nonumber\\
    &&+(J_{00\mu'\nu'}^{rr'}-J_{0000}^{rr'} \delta_{\mu'\nu'})b_{r'}^{\mu'\dagger} b_{r'}^{\nu'}],
    \label{eq9}
\end{eqnarray}
where the index $\mu,\nu,\mu',\nu'\neq0$ and will be summed up when appear twice in a single term, and the tilde~ $\tilde{}$ ~on $J_{\mu\nu\mu'\nu'}^{rr'}$ and $b_r^\mu$, which denotes the expressions after the unitary transformation that minimizes the mean field variational energy, is omitted for simplicity.

Now Eq.~(\ref{eq9}) is a free bosonic Hamiltonian and can be solved by performing the Fourier transformation,
\begin{equation}
 b_k^\mu={1\over \sqrt{L}}\sum_r b_r^\mu e^{i{\bf k}\cdot{\bf r}},
\end{equation}
with $L$ the lattice number of the system. It leads to,
\begin{eqnarray}
H&=&\sum_k\psi_k^\dagger h(k) \psi_k,\nonumber
  \\
  \psi_k&=&(b_k^1,...,b_k^{M(n-1)},b_{-k}^{1\dagger},...,b_{-k}^{M(n-1)\dagger})^T,
  \label{eq10}
\end{eqnarray}
where $M$ is the size of magnetic cell.
There are two diagonalization methods for a bosonic Hamiltonian as proposed by White\cite{White1965} and Colpa\cite{Colpa1978}. After diagonalization, we get the spin-wave dispersion $\epsilon_\mu(k)$ as expressed by,
\begin{eqnarray}
  H&=&\sum_{\mu=1}^{M(n-1)}\epsilon_\mu(k)\gamma_k^{\mu \dagger}\gamma_k^\mu, \nonumber
  \\
  \gamma_k^\mu&=&T_{\mu'}^\mu b_k^{\mu'},
  \label{eq11}
\end{eqnarray}
with $T_{\mu'}^\mu$ the element of the matrix used to diagonalize the Hamiltonian.
As noted, the SU($N$) spin-wave theory includes not only the dipole-dipole correlations, but also the multipole-multipole correlations. In general, the correlation function of two SU($N$) generators can be written by,
%{\footnotesize
\begin{eqnarray}
  S^{\mu\nu\mu'\nu'}(k,\omega)&=&{1\over 2M(n-1)}\int dt e^{-i\omega t} \nonumber \\
  & &\times\Sigma_{r,r'}e^{i{\bf k}\cdot({\bf r}-{\bf r'})} \langle S_r^{\mu\nu}S_{r'}^{\mu'\nu'}(t)\rangle.\label{eq13}
\end{eqnarray}
%}
As same as the SU($2$) linear spin-wave theory, only the quadratic forms of the dynamical part of correlation functions are calculated. Therefore, the correlation function is expanded in $\langle b^{\mu\dagger}b^\mu\rangle$, which describes the probability to excite one of bosonic excitations. It is clear that there are $M(n-1)$ spin-wave modes.

\section{SU($4$) antiferromagnetism}\label{su4}
As an example, we first calculate the spin-wave spectrum for the SU($4$) antiferromagnetic model in a square lattice. The model can be generated from the generic one-band Hubbard model loaded with spin-$3/2$ fermions. Due to Pauli¡¯s exclusion principle, the wave functions of two on-site fermions have to be antisymmetric. The total spin of two on-site spin-$3/2$ fermions can only be either singlet ($S =0$) or quintet ($S =2$). So the effective model at quarter-filling will have only two exchange channels, and the spin singlet channel results in the SU($4$) antiferromagnetic Hamiltonian:
\begin{equation}
  H=J\sum_{\langle i,j\rangle}\left[\sum_{1\leq a<b\leq5}\Gamma_i^{ab}\Gamma_j^{ab}-\sum_{a=1}^5\Gamma_i^a\Gamma_j^a\right],
  \label{eq14}
\end{equation}
where $\Gamma^a$ are Dirac matrices which form Clifford algebra, $\{\Gamma^a,\Gamma^b\}=2\delta^{ab}$ and $\Gamma^{ab}=\left[\Gamma^a,\Gamma^b\right]/(2i)$. Specifically, the five Dirac matrices can be expressed as tensor products of tow Pauli spin-$1/2$ matrices $(\sigma^\alpha,\tau^\beta)$, or represented by symmetric bilinear combinations of the components of a spin-$3/2$ operator, $S^x,S^y,S^z$:
\begin{align*}
  &\Gamma^1=\sigma^z\tau^y={1\over\sqrt{3}}\left\{S^y,S^z\right\}, \\ &\Gamma^2=\sigma^z\tau^x={1\over\sqrt{3}}\left\{S^x,S^z\right\}, \\
  &\Gamma^3=\sigma^y\tau^0={1\over\sqrt{3}}\left\{S^x,S^y\right\}, \\ &\Gamma^4=\sigma^x\tau^0={1\over\sqrt{3}}\left[(S^x)^2-(S^y)^2\right],\\
  &\Gamma^5=\sigma^z\tau^z=(S^z)^2-{5\over4}.
\end{align*}
First of all, the spin exchange Hamiltonian stems from a SU($2$) symmetrical one-band Hubbard model with spin-$3/2$ fermions, so it has the genetic SU($2$) symmetry. Also, all $15$ Gamma operators together span the SU($4$) algebra. Among them, the $10$ $\Gamma^{ab}$ operators are SO($5$) anti-symmetric tensors, while the five $\Gamma^a$ are SO($5$) vectors. Thus the Hamiltonian Eq.~(\ref{eq14}) obviously possesses SO($5$) symmetry. Moreover it also has a hidden SU($4$) symmetry in the bipartite lattice\cite{Wu2003}. We can define a particle-hole transformation $b^\mu\rightarrow\mathcal{J}b^{\mu\dagger}$ with an antisymmetric matrix $\mathcal{J}=i\sigma^x\tau^y$. With this operation, the fundamental representation transforms to a conjugate representation where $\Gamma^{ab*}=\Gamma^{ab}$ and $\Gamma^{a*}=-\Gamma^{a}$. If transforming all $B$ sublattices into the conjugate representation, then we have,
\begin{equation}
  H=J\sum_{\langle i,j\rangle}\left[\sum_{1\leq a<b\leq5}\Gamma_i^{ab*}\Gamma_j^{ab}+\sum_{a=1}^5\Gamma_i^{a*}\Gamma_j^a\right].
  \label{eq15}
\end{equation}
One should note that Eq.~(\ref{eq14}) is invariant under SU($4$) rotations and conjugate rotations on sublattices $A$ and $B$, respectively, rather than under uniform SU($4$) transformations.

\begin{figure}
  \centering
  \includegraphics[width=0.4\textwidth]{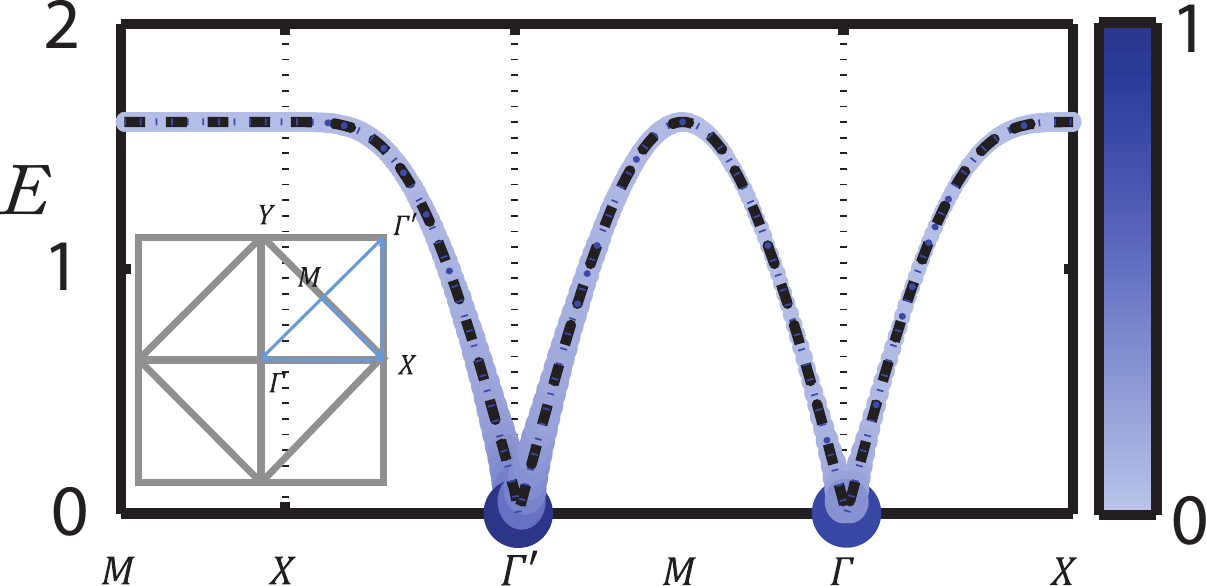}
  \caption{\label{fig1}(Color online) Spin waves of the SU($4$) antiferromagnetic model in a square lattice along high symmetry directions. The dashed lines denote the dispersions, and the size and color of the marks indicate the intensity of the quadrupolar-quadrupolar correlation function.}
\end{figure}

In a square lattice, the SU($4$) linear spin wave theory shows a long-range Neel order which is consistent with the quantum Monte Carlo simulations\cite{Harada2003}. There are three local order parameters of SU($4$) Neel order in the square lattice: $\left(\Gamma^{12},\Gamma^{34},\Gamma^5\right)=\left((-1)^{x+y}m,(-1)^{x+y}m,m\right)$. In the case of spin-$3/2$, they can be expanded in multipole orders as defined in Eq.~(\ref{eq4}):
\begin{align*}
  \Gamma^{12} &= {2\over \sqrt{5}}(2M_{1,0}-M_{3,0}), \\
  \Gamma^{34} &= {2\over \sqrt{5}}(M_{1,0}+2M_{3,0}), \\
  \Gamma^5 &=2M_{2,0}.
\end{align*}
Therefore, we choose to calculate a quadrupolar-quadrupolar correlation function along high symmetry directions,
\begin{equation*}
M_2(k,\omega)\propto\sum_{r,r'}e^{i{\bf k}\cdot({\bf r}-{\bf r'})}\int dt e^{-i\omega t} \left\langle \sum_m M_{2,m}({\bf r})M_{2,m}^{\dagger}({\bf r},t)\right\rangle,
\end{equation*}
The numerical results are shown in Fig.~\ref{fig1}.
The Goldstone manifold is CP($3$) = U($4$)$/[$U(1)$\bigotimes $U($3$)$]$ with $6$ branches of spin waves, which are degenerated and look like the dispersion of the SU($2$) antiferromagnetic spin waves in a square lattice. However,
the quadrupolar-quadrupolar correlation exhibits a noticeable intensity at the $\Gamma=(0,0)$ point as shown in Fig.~\ref{fig1}. It is in sharp contrast to the behavior of the antiferromagnetic spin-spin correlation, which vanishes at that point.

\section{SU($N$) spin wave study of TMOs}\label{tmo}

As we know, most of TMOs have a magnetic ordered ground state. Considering that these magnetic ordered states can be described by isospins which are the entangled states of spin and orbital degrees of freedom, we use the SU($N$) spin wave theory to investigate excitations from the ordered state.
%In contrast to the traditional SU($2$) spin wave theory, it is a more accurate and detailed way to describe the system.
%Now we attempt to apply the SU($N$) spin wave theory to the TMOs.
%It is known that $d$ orbitals have five orbitals and more than one electrons is active, so it is difficult to include all the degrees of freedom in simulation.
We will first present a general method to derive the effective exchange model from an electron model in the strong interaction limit.
We consider the multi-band Hubbard model which is suitable to describe properties of TMOs,
\begin{equation}
  H=\sum_{\langle ij\rangle,\alpha\alpha'}t_{\alpha\alpha'}^{ij} c^{\dagger}_{i\alpha}c_{j\alpha'}+\sum_iH_i.
  \nonumber
\end{equation}
Here the first term is hopping terms with $t_{\alpha\alpha'}^{ij}$ the element of hopping integrals, and $\alpha$ indicates all the local degrees of freedom, such as orbitals and spins. $H_i$ are the local interactions which include the multi-band Hubbard term $V_i$, SOC $O_i$, and local potential field $W_i$,
\begin{align}
V_i&=
\frac{1}{2}\sum_{mm'nn'}\sum_{\sigma\tau\mu\nu}\delta_{\sigma\nu}\delta_{\tau\mu}
\{U\delta_{m=m'=n=n'} (1-\delta_{\sigma\tau}) \nonumber \\
&+ U^{\prime}\delta_{mn'}\delta_{m'n}(1-\delta_{mm'}) + J_{h}\delta_{mn}\delta_{m'n'}(1-\delta_{mm'}) \nonumber \\
&+ J^{\prime}\delta_{mm'}\delta_{mn'}(1-\delta_{mn})(1-\delta_{\sigma\tau})\} \nonumber \\
&~~~~~~~~~~~~~~~~~~~~~~~~~~~~~\cdot c^\dag_{im\sigma}c^\dag_{im'\tau}c_{in\mu}c_{in'\nu},
\label{eq:Hu}\\
O_i&=\lambda \bm{S}_i\cdot \bm{L}_i,\label{eq:SOC}\\
W_i&= \sum_{\alpha\beta} w_{i\alpha\beta}c_{i\alpha}^\dagger c_{i\beta}.\label{eq:SOC}
\end{align}
where $U$ ($U^{\prime}$) is the intra-orbital (inter-orbital) Coulomb interaction, $J_{h}$ and $J^{\prime}$ are the Hund's coupling and the pairing hopping, respectively. In this paper, we employ $U=U^\prime+2J_{h}$ and $J^\prime=J_{h}$ as used usually.

By means of the perturbation theory, we treat the hopping terms as the perturbation in the strong interaction limit and obtain the effective exchange model which can be generally written as,
\begin{eqnarray}
  H_{\rm eff}&=&\sum_iP_i^0H_iP_i^0+\sum_{\langle i,j\rangle}\left[H_{i\rightarrow j}+H_{j\rightarrow i}\right],
  \label{eq16}\\
  H_{i\rightarrow j}&=&\sum_{ \footnotesize{\begin{array}{c}
                               (lre)\\
                               \alpha\alpha'\beta\beta
                             \end{array}}
  }{1\over \Delta_{lre}}t_{\alpha'\alpha}^{\langle ij\rangle}\left[s_i^{\alpha'\beta'}\right]_{(lre)}t_{\beta\beta'}^{\langle ji\rangle}\left[\tilde{s}_j^{\beta\alpha}\right]_{(lre)}.\label{eq17}
\end{eqnarray}
The first term in Eq.~(\ref{eq16}) is the zero and first order perturbation term, and the second is the second order perturbation term accounting for the virtual hoppings of electrons contributing to spin exchanges. $P_i^0$ is the operator projecting the Hamiltonian $H_i$ into its low-energy subspace.
$s_i^{\alpha\beta}=c^{\dagger}_{i\alpha}c_{i\beta}$ and $\tilde{s}_i^{\beta\alpha}=c_{i\alpha} c_{i\beta}^{\dagger}$ are SU($N$) generators and their conjugate representation, respectively. $(lre)$ denotes various scattering channels related to the virtual processes from a low energy state $|\psi_r\rangle=\prod_i|r_i\rangle$ to a high one $|\psi_e\rangle=\prod_i|e_i\rangle$, and back to the low one $|\psi_l\rangle=\prod_i|l_i\rangle$, where $\prod_i|r_i\rangle$ is the eigenstate of Hamiltonian $\sum_iH_i$.
$1/ \Delta_{lre}= 1/2(E_{li}+E_{lj}-E_{ei}-E_{ej})+ 1/2(E_{ri}+E_{rj}-E_{ei}-E_{ej})$, in which $E_{mi}$ ($m=l,e,r$) is the eigenenergy of the local state $|m_i\rangle$ on the site $i$. $[~]_{(lre)}$ indicates a special representation of $s_i^{\alpha\beta}$ and $\tilde{s}_j^{\beta\alpha}$ in the states $(|l_i\rangle,|r_i\rangle,|e_i\rangle)$
\begin{eqnarray*}
  \left[s_i^{\alpha\beta}\right]_{(lre)} &=& |l_i\rangle\langle l_i|c_i^{\alpha\dagger}  |e_i\rangle\langle e_i| c_i^\beta |r_i\rangle\langle r_i|,\nonumber\\
  &=&\langle l_i|c_i^{\alpha\dagger}  |e_i\rangle\langle e_i| c_i^\beta |r_i\rangle S_i^{l_ir_i},\\
  \left[\tilde{s}_i^{\beta\alpha}\right]_{(lre)} &=& |l_i\rangle\langle l_i|c_i^{\alpha}  |e_i\rangle\langle e_i| c_i^{\beta\dagger} |r_i\rangle\langle r_i|,\nonumber\\
  &=&\langle l_i|c_i^{\alpha}  |e_i\rangle\langle e_i| c_i^{\beta\dagger} |r_i\rangle S_i^{l_ir_i},\\
\end{eqnarray*}
where $S_i^{l_ir_i}=|l_i\rangle\langle r_i|$ is the SU($N$) generator in the fundamental representation defined on the low-energy space of $H_i$.
%We note that $s_i^{\alpha\beta}$ satisfies the SU($N$) commutation relationship.
%and Eq.~(\ref{eq17}) is SU($N$) antiferromagnetic Hamiltonian if disregarding the $[~]_{(lre)}$ representation and anisotropic hopping integrals $t_{\alpha'\alpha}^{\langle ij\rangle}$. %In the $[~]_{(lre)}$ representation, the SU($N$) commutation relations of $s_i^{\alpha\beta}$ may degenerate to a plainer commutation relations and
We note the symmetry of Hamiltonian Eq.~(\ref{eq17}) is related to the symmetry of  $(|l_i\rangle,|r_i\rangle,|e_i\rangle)$ and $t_{\alpha'\alpha}^{ij}$, which are determined by the symmetry of the crystal structure.
%We would not expand more on Eq.~(\ref{eq17}) and its additional properties, which will be derived and explained in detail in our another paper\cite{prepared}.
Now with Eq.~(\ref{eq16}), we will carry out the SU($N$) spin wave calculation.

\subsection{Three band Hubbard model with an SOC on the hexagon lattice}

As an illustration of the application of the SU($N$) spin wave theory, let us first consider a simple three band Hubbard model with one spin-$1/2$ particle per site and SOC, $-\lambda \vec{s}\cdot\vec{l}$ (The minus sign is due to that $l$ is a mirror angular momentum) on the hexagon lattice. The Hubbard term presents SU($2$) and SO($3$) symmetry with $U=U'+2J_h$. Focusing on the effect of Hund's coupling and SOC, we suppose a simply isotripic hopping term, $t_{\alpha'\alpha}^{ij}=t\delta_{\alpha'\alpha}$ only among the nearest neighbours.

If SOC is absent, its effective exchange model is comparatively explicit. Because the wave functions of two on-site fermions have to be antisymmetric, there are only three exchange channels. The initial and final low energy states are singly occupied states with zero energy, and three intermediate states which are vacuum states on one site and doubly occupied states on the other site with 1) total spins are $S=1$, total orbital momentums $L=1$ and $\Delta_{lre}=-U+3J_h$, 2) total spins are $S=0$, total orbital momentums $L=2$ and $\Delta_{lre}=-U+J_h$ and 3) total spins are $S=0$, total orbital momentums $L=0$ and $\Delta_{lre}=-U-2J_h$.
However, when SOC is comparable to the Hubbard term, $\lambda\sim U$, there will be $20=2\times5\times2$ channels due to the interplay of the SOC and Hund's coupling: two kinds of initial and final states with energy $\lambda/2$ and $-\lambda$ respectively, and five kinds of intermediate states with energy $U-3J_h-\lambda/2,(2U-J_h-\lambda\pm\sqrt{25J_h^2+10J_h\lambda+9\lambda^2})/2$ and $(4U-8J_h+\lambda\pm\sqrt{16J_h^2+8J_h\lambda+9\lambda^2})/4$. Substituting $\Delta_{lre}$ with the corresponding $(|l_i\rangle,|r_i\rangle,|e_i\rangle)$ and $t_{\alpha'\alpha}^{ij}$ into Eq.~(\ref{eq17}), we can easily obtain the exchange model numerically.

If $J_h=0,\lambda=0$, $H_i$ has SU($6$) symmetry, so does $(|l_i\rangle,|r_i\rangle,|e_i\rangle)$ and $t_{\alpha'\alpha}^{ij}$, but the symmetry of eigenstates will be broken into SU($2$) by either SOC or Hund's coupling. Furthermore, when $t_{\alpha'\alpha}^{ij}$ is SU($2$) symmetrical, the effective Hamiltonian must be SU($2$) symmetrical too. If $\lambda\gg J_h$, only the lowest energy channel is active. In this case, the Hamiltonian can be further approximated to be an effective isospin-$1/2$ model. However, the Hund's coupling will lower the energy of the spin parallelling states of two electrons, while the SOC will lower the energy of single electron $J_{\rm eff}=s-l=1/2$ states. This would influence the validity of the isospin $J_{\rm eff}=1/2$ model. Therefore, we intend to take both $\lambda$ and $J_h$ into account to examine the SU($6$) spin-wave spectrum of the system.

\begin{figure}
  \centering
  \includegraphics[width=0.3\textwidth]{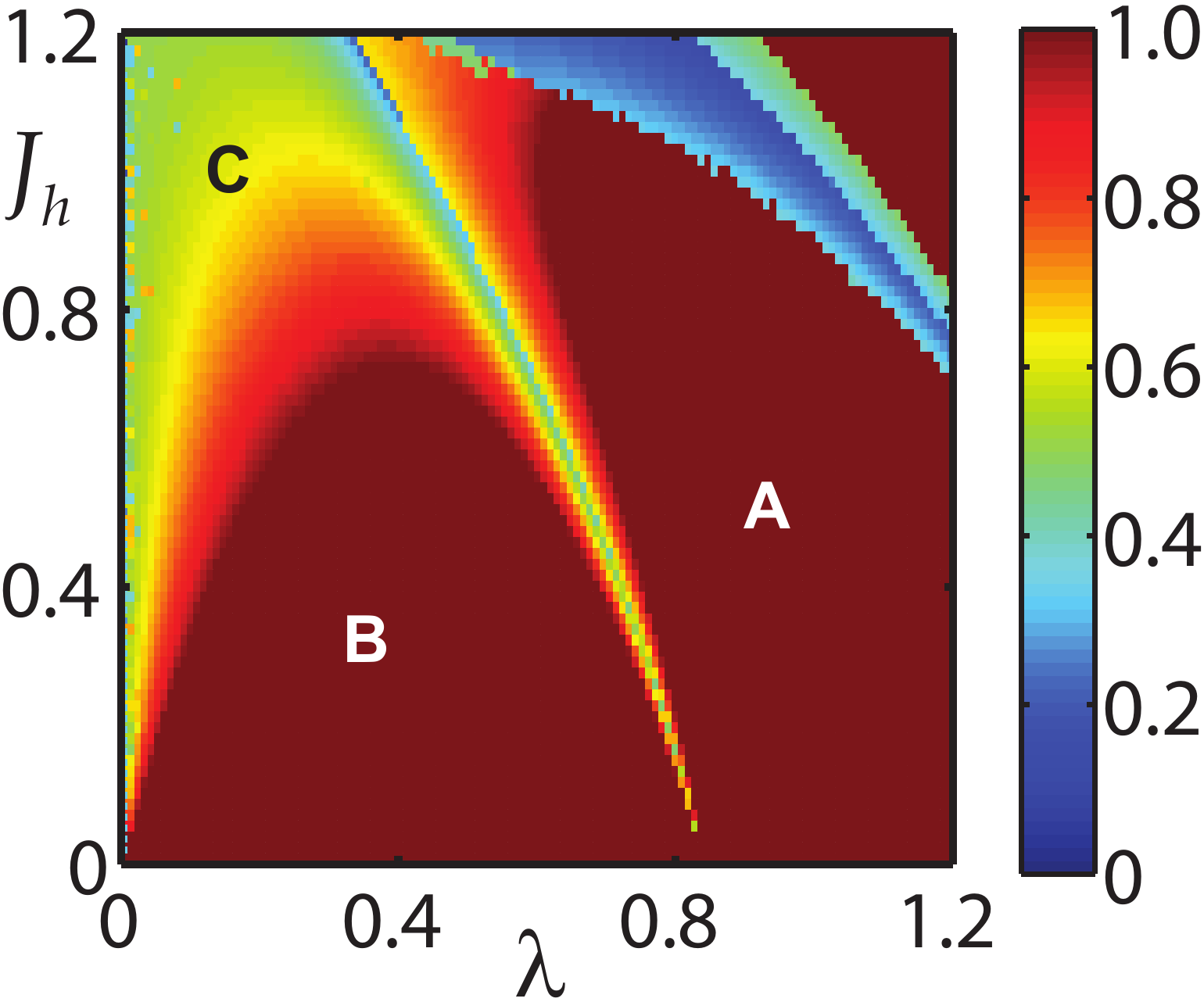}
  \caption{\label{fig2}(Color online) Weights of the $J_{\rm eff}=1/2$ states in ground states vary with $\lambda$ and $J_h$, calculated based on the three band Hubbard model with an SOC on the hexagon lattice. The intra-orbital Coulomb interaction is $U=5.0$ }
\end{figure}

Firstly, the local mean field theory suggests a magnetic cell with two sites, so we suppose the local mean field wave function in two sublattices of the hexagon lattice are $|T_A\rangle$ and $|T_B\rangle$. In order to verify the validity of the isospin $J_{\rm eff}=1/2$ model, we calculate the weight $(\langle J_{\rm eff}=1/2|T_A\rangle$+$\langle J_{\rm eff}=1/2|T_B\rangle)/2$ of $J_{\rm eff}=1/2$ states in ground states as shown in Fig.~\ref{fig2}. We use the hopping term $t=1$ as unit, set $U=5.0$ and change $\lambda$ and $J_h$ from $0$ to $1.2$. There are roughly three regions: \textbf{A}. Rightside region in which the ground states are dominated by $J_{\rm eff}=1/2$ states; \textbf{B}. A bump in the area of small $\lambda$ and $J_h$ where ground states are also dominated by $J_{\rm eff}=1/2$ states; \textbf{C}. $J_h$ is so large that the ground states are mixed by the $J_{\rm eff}=3/2$ states. The blue discontinuous region on the right top is due to the divergence of the second order perturbation, which means the SOC gap is comparable to the Hubbard gap. Thus the low energy physics can certainly be described by the $J_{\rm eff}=1/2$ doublet in the region beyond this discontinuous region (where the SOC is dominated).

\begin{figure}
  \centering
  \includegraphics[width=0.3\textwidth]{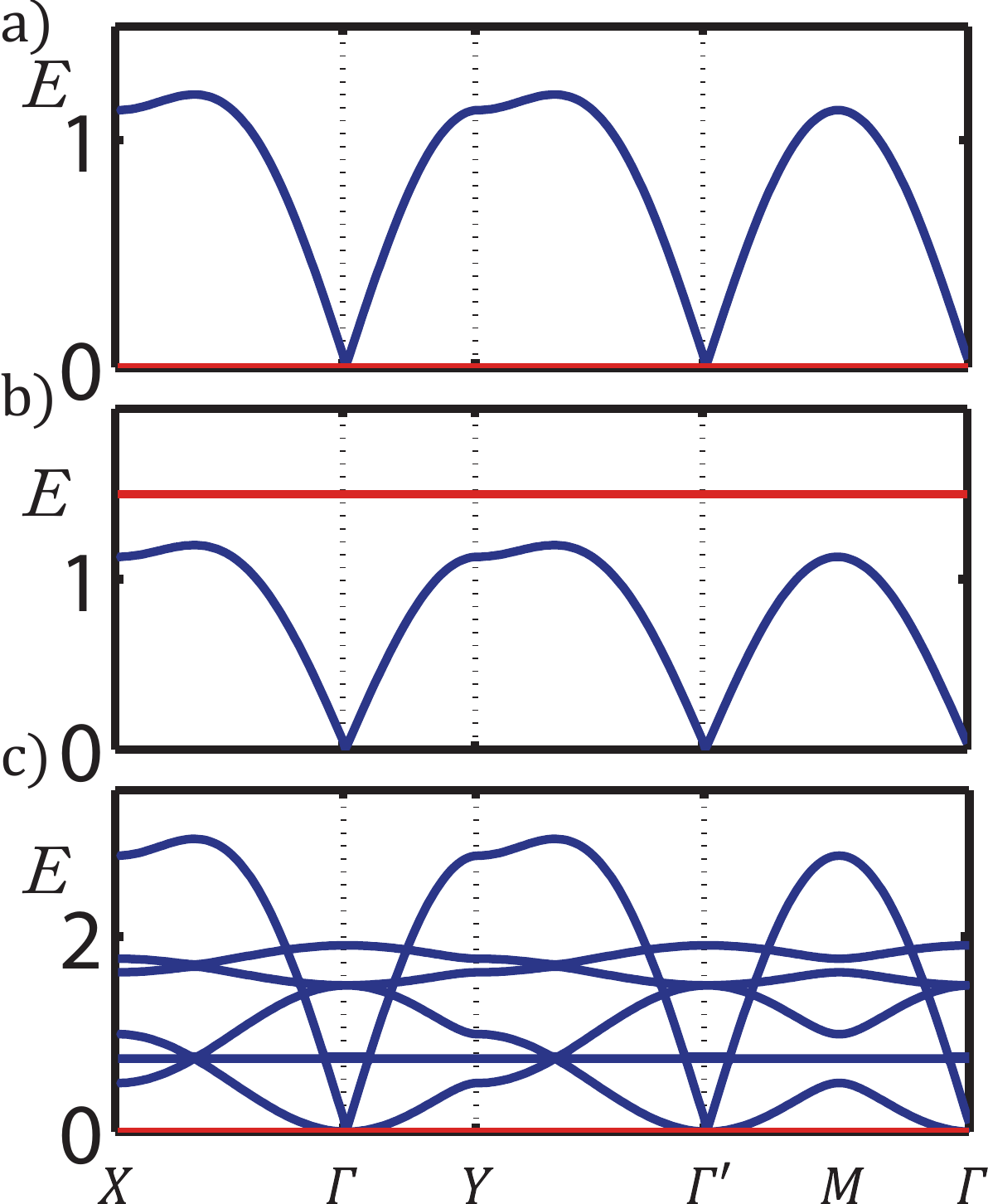}
  \caption{\label{fig11}(Color online) Spin waves of three band Hubbard model on the hexagon lattice with SOC $\lambda$ and Hund's coupling $J_h$, which are: a) $\lambda=0,J_h=0$, b) $\lambda>0,J_h=0$ and c) $\lambda=0,J_h>0$.}
\end{figure}

Let us first consider some extreme situations. The calculated dispersions for spin excitations in three band Hubbard model based on the spin wave theory for several cases are shown in Fig.~\ref{fig11}. When $J_h=0$ and $\lambda=0$, there are highly degenerated zero energy spin waves suggesting that the magnetic order are unstable, as shown in Fig.~\ref{fig11} a). This is because the ground state is the SU($6$) plaquette state\cite{Nataf2016,Zhao2012} in this situation, where SU($6$) spins form local singlets on a hexagon plaquette. There is no long-range ordering on which the SU($N$) spin wave theory is based, so the spin wave theory fails in this case. As $\lambda$ increases, the zero energy spin waves are lifted [see Fig.~\ref{fig11} b)], and the system approaches ordered phases because the fluctuations become weak gradually as the system departs the SU($6$) symmetry due to SOC. On the other hand, there is a ferromagnetic-like spin wave emerging when turning on the Hund's coupling $J_h$ instead of SOC $\lambda$, as shown in Fig.~\ref{fig11} c). However, there is still some zero energy degeneracies. Thus, the ground state may be still an SU($6$) plaquette state or some RVB states.
\begin{figure}
  \centering
  \includegraphics[width=0.46\textwidth]{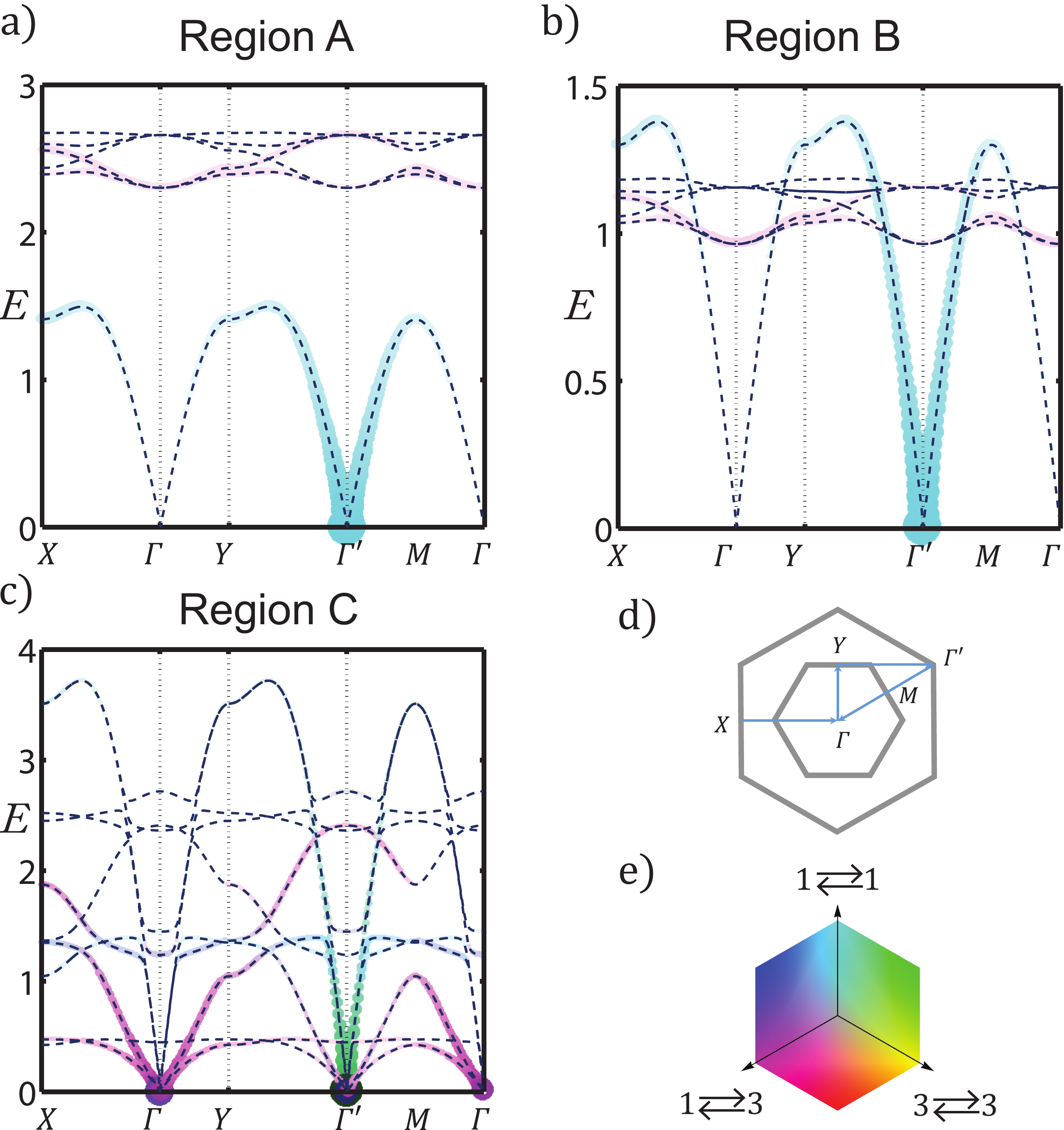}
  \caption{\label{fig3}(Color online) Spin waves with paremeters: a) $\lambda=0.9,J_h=0.6$, b) $\lambda=0.4,J_h=0.4$ and c) $\lambda=0.2,J_h=1.1$. The dashed lines denote dispersions. The size and saturation of makers indicate the intensity of correlation function, and three different channels are indicated by three different colors. e)Reciprocal lattices and high symmetry directions of a hexagon lattice. d)The legend indicating the compositions of the correlation function.}
\end{figure}

Then, we study the correlation functions in three regions \textbf{A},\textbf{B} and \textbf{C}, respectively. In the dipole-dipole approximation, the correlation function consists of three parts of contributions: spin flippings within either $J_{\rm eff}=1/2$ or $3/2$ states and spin flippings across the $J_{\rm eff}=1/2$ and $J_{\rm eff}=3/2$ states, which are denoted by $1\rightleftarrows1$, $3\rightleftarrows3$ and $1\rightleftarrows3$, respectively. In Figs~\ref{fig3} a)-c), we present the dispersions of spin waves denoted by the dashed lines and intensities of the correlation functions indicated by the saturation of three different colors and size of markers. The colors will mix as shown by the legend in Fig.~\ref{fig3} e), when spin wave excitations includes more than two types of contributions.
In region \textbf{A}, the result suggests an antiferromagnetic-like spin wave at low energies, which is linear around $\Gamma$ point and the intensity diverges at $\Gamma'$ but vanishes at $\Gamma$ point, and a ferromagnetic-like spin wave at high energies above $2$, which is parabolic around $\Gamma$ point and the intensity is higher at $\Gamma$ than $\Gamma'$ point. At the meantime, the result calculated by using the local mean field theory shows the system has a $J_{\rm eff}=1/2$ antiferromagnetic ordered ground state, confirming that the excitations at low energies are indeed antiferromagnetic spin waves. As shown by the cyan-blue color in Fig.~\ref{fig3} a), these low-energies excitations comes basically from spin flippings within the $J_{\rm eff}=1/2$ states, so the low-energy physics in region \textbf{A} is dominated by isospin-$1/2$ states. Furthermore, the excitations arising from the spin flippings across the $J_{\rm eff}=1/2$ and $J_{\rm eff}=3/2$ states as denoted by the magenta color are far beyond the low-energy excitations due to the sufficiently large SOC. Thus, we arrive at the conclusion that an effective isospin Heisenberg model can depict the low-energy physics in region \textbf{A}, which is also consistent with the calculation of weights of $J_{\rm eff}=1/2$ states in ground states as shown in Fig.~\ref{fig2}.
When the SOC is decreased, we will enter gradually into region \textbf{B}. In this progress, the gap between the low-energy antiferromagnetic spin wave and the high-energy ferromagnetic spin wave decreases gradually. However, as long as $J_h$ is not large enough, although the dispersion of ferromagnetic spin waves overlaps with the low energy one, the two spin waves do not entangle each other, as indicated by Fig.~\ref{fig3} b) where the colors representing two different kinds of spin waves do not mix. Thus, apart from the effective isospin Heisenberg terms in the Hamiltonian, which describes the antiferromagnetic spin waves, there have to be another term to describe the ferromagnetic spin waves at least.
%In region \textbf{A} and \textbf{B} there are two types of well-defined dispersions. One is an antiferromagnetic spin wave, which is linear around $\Gamma$ point and the intensity is diverging at $\Gamma'$ but vanishing at $\Gamma$. The other is an ferromagnetic spin wave, which is parabolic around $\Gamma$ point and the intensity is higher at $\Gamma$. In both the two region, the ground state is $J_{\rm eff}=1/2$ antiferromagnetic ordered, so the cyan-blue antiferromagnetic spin wave is the spin flipping of $J_{\rm eff}=1/2$. For the excitations of jumping from $J_{\rm eff}=1/2$ to $J_{\rm eff}=3/2$, the ground state appears to be a ferromagnetic-like state, which is the reason why the magenta spin wave looks like a ferromagnetic one. In region \textbf{A}, the two spin wave dispersions are isolated, so we believe a $J_{\rm eff}=1/2$ isospin model is enough to capture all the low energy physics. In region \textbf{B}, although two spin waves do not entangle each other, the low energy physics include two parts that suggest the Hamiltonian containing at least two different kinds of terms.
Starting from region \textbf{B}, one can increase $J_h$ to enter into region \textbf{C}. In this region, the antiferromagnetic and ferromagnetic spin waves are entangled, so that there is no well-defined antiferromagnetic-like spin waves or ferromagnetic-like spin waves, and the local test wave functions of ground state in two different sublattices are not completely orthogonal, namely $\langle T_A|T_B\rangle\approx0.016$. Because the ground state consists of both $ J_{\rm eff}=1/2$ and $ J_{\rm eff}=3/2$ states now, the multipolar orders are inevitable to be taken into account. Its dipolar order parameters $\langle J_{\rm eff}^\alpha\rangle$ are almost antiferromagnetic, but quadrupolar order parameters $\langle J_{\rm eff}^\alpha J_{\rm eff}^\beta+J_{\rm eff}^\beta J_{\rm eff}^\alpha\rangle$ are ferromagnetic. In this case, all degrees of freedom have to be taken into account and there is no so-called isospin effective Hamiltonian, so the SU($N$) spin wave theory rather than the traditional SU($2$) one is applicable.

\subsection{$\alpha$-RuCl$_3$ and Sr$_2$IrO$_4$ }

In this subsection, we will use the SU($N$) spin-wave theory to study spin dynamics in $\alpha$-RuCl$_3$ and Sr$_2$IrO$_4$. Both $\alpha$-RuCl$_3$ and Sr$_2$IrO$_4$ have a $d^5$ configuration and have an octahedral crystal field. Their differences are that the active electrons residing in $4d$ orbitals of Ru has a smaller SOC than that in $5d$ of Ir, and $\alpha$-RuCl$_3$ is a honeycomb lattice while Sr$_2$IrO$_4$ is a square lattice.

\begin{figure}
  \centering
  \includegraphics[width=0.40\textwidth]{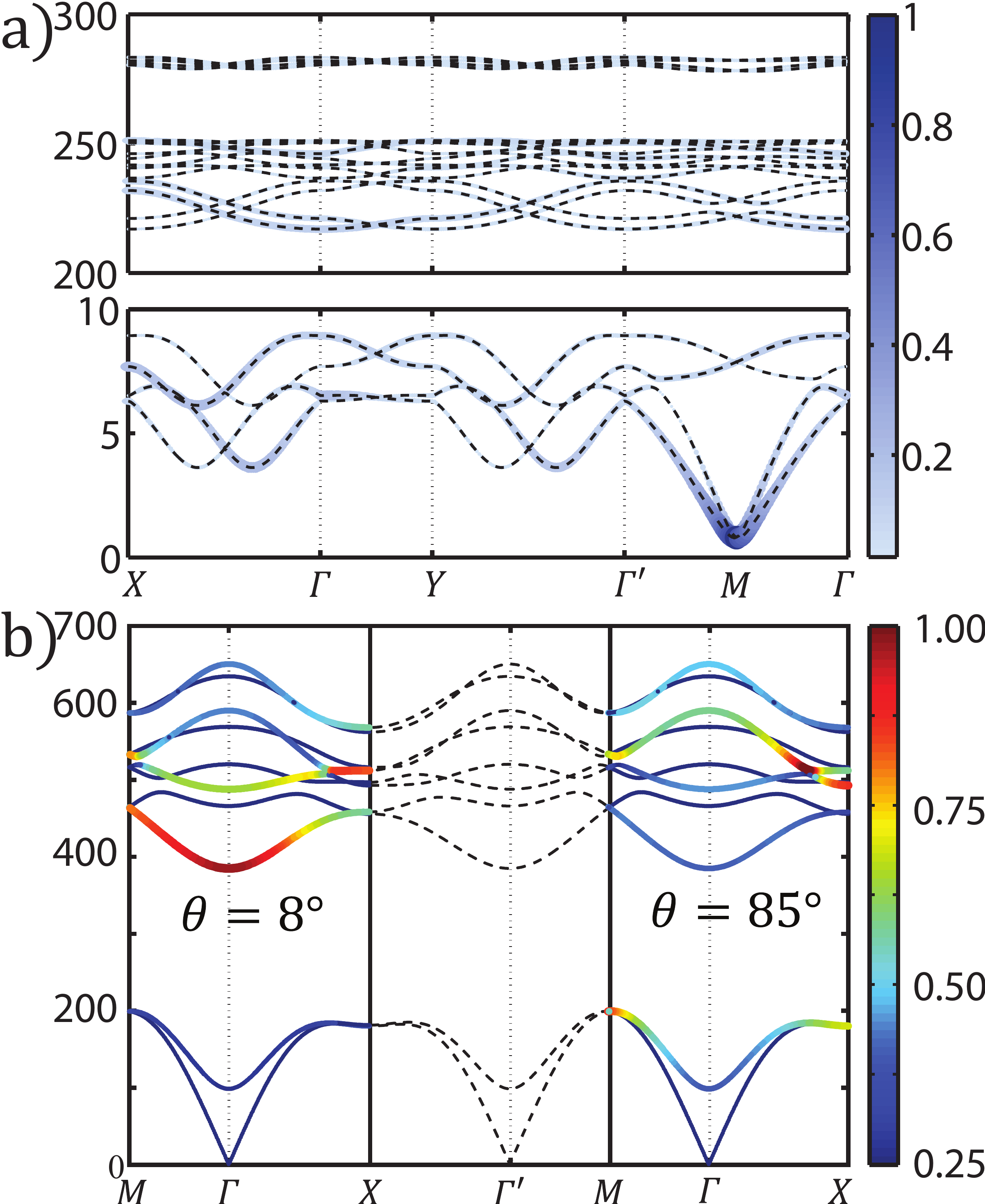}
  \caption{\label{fig4}(Color online) Spin-spin correlation functions for $\alpha$-RuCl$_3$ a), and correlation functions of RIXS operators\cite{Luo1993} for Sr$_2$IrO$_4$ b) along the high-symmetry lines, calculated by the SU($N$) spin-wave theory.}
\end{figure}

$\alpha$-RuCl$_3$ has a layered crystal structure with $\mathrm{Ru}^{3+}$ forming the honeycomb lattice layers and the energy bands near the Fermi level are dominated by the $d$ orbitals of Ru. We consider a five band tight-binding model with five electrons per site and the on-site crystal fields to describe the $4d^5$ configuration of $\mathrm{Ru}^{3+}$. The tight-binding parameters include the nearest-, next-nearest- and third-nearest-neighbour hopping integrals, which are obtained by fitting to the energy-band dispersions calculated by the first principle calculations and given in our previous paper Ref.~[\onlinecite{Wang2017}]. We take $U=2.7~{\rm eV},J_h=0.13U, U'=U-2J_h,~{\rm and}~\lambda=0.14~{\rm eV}$\cite{PhysRevLett.117.126403,PhysRevB.93.075144,Banerjee2016,PhysRevB.93.214431,PhysRevB.91.241110,Wang2017} in the following calculations. Then, an effective exchange model is obtained numerically according to Eq.~(\ref{eq16}). Due to the large crystal field potential on the $e_g$ orbitals, there are isolated six lowest energy states, onto which we will project the initial and final states. Using the local mean field theory and the SU($N$) Linear spin-wave approximation introduced in Sec.~\ref{spinwave}, we investigate numerically the magnetic ground state and spin dynamics. Numerical results
show that the magnetic ground state has a zigzag type order of which the magnetic unit cell contains four sites (two cells), in agreement with experiments in $\alpha$-RuCl$_3$\cite{Sears2015,Banerjee2016,Ran2017}
The spin-spin correlation functions calculated by Eq.~(\ref{eq13}) is shown in Fig.~\ref{fig4} a).
Below $30$ meV, four zigzag spin waves are evident, and the other sixteen excitations around $200$ meV come from the spin-orbital excitations across the $J_{\rm eff}=1/2$ and $J_{\rm eff}=3/2$ states.
Though there is long-range zigzag spin order, the results in Fig.~\ref{fig4} a) show that the low-energy spin waves have a gap of about $2$ meV at $M$ point and the spin-spin correlation function has a maximum magnitude also at $M$ point. These results are consistent with the recent experiments of inelastic neutron scatterings on $\alpha$-RuCl$_3$\cite{Banerjee2016,Ran2017}.
On the other hand, the gap between the zigzag spin waves and the spin-orbital excitations is of about $210$ meV, thus suggests that the low energy physics of $\alpha$-RuCl$_3$ could be captured by an effective isospin-$1/2$ model. We have found in our previous paper Ref.~[\onlinecite{Wang2017}] that the minimum effective isospin-$1/2$ model is the K-$\Gamma$ model containing a ferromagnetic nearest-neighbor Kitaev interaction (K) and a nearest-neighbor off-diagonal exchange interaction ($\Gamma$).

%Moreover the little ingredients of $e_g$ orbitals in the ground state plays an important role in the determination of the sign of Kitaev interaction.

%The concrete parameters and process will be clarified in the reference~[\onlinecite{Wang2017}], and the results will be discussed in detail as well.

%\begin{equation}
%  \chi^{RPA}(q,\omega)=[1-\chi^0(q,\omega)\mathcal{J}(q)]^{-1}\chi^0(q,\omega)
%\end{equation}
%
%\begin{equation}
%  \mathcal{J}(q)=\left(
%    \begin{array}{ccc}
%      0 & -2\sqrt{3}w_qJ_{\pm\pm} & -\sqrt{3}w_qJ_{z\pm} \\
%     -2\sqrt{3}w_qJ_{\pm\pm} & 0 & (u_q-v_q)J_{z\pm} \\
%      -\sqrt{3}w_qJ_{z\pm} &(u_q-v_q)J_{z\pm} & 0 \\
%    \end{array}
%  \right)
%\end{equation}

Now let us turn to Sr$_2$IrO$_4$. We start our investigations from a three band Hubbard model with a single hole per site to fit the band dispersion around the Fermi level\cite{Watanabe2010,Wang2015}, and choose $U=3.6~{\rm eV},J_h=0.18U,~{\rm and}~\lambda=0.37~{\rm eV}$ in the calculation.
Because iridium is a strong absorber of neutrons, it is more useful to calculate the resonant inelastic X-ray scattering (RIXS) spectrum for the purpose of a comparison with experiments.
RIXS involves a second order process that includes an absorption and an emission of a photon. In the fast collision approximation, the direct RIXS spectrum is proportional to the correlation function of spin-orbital moment operators\cite{Luo1993}. Due to the two scattering progresses (absorption and emission), the total angular momentum of spin-orbital moment operators is equal to the coupling of two $l=1$ angular momenta (angular momentum exchange of the two scatterings is one in the dipole limit). Thus, there exists multipole-multipole correlations in RIXS besides the usual dipole-dipole correlations. It is known that the RIXS spectrum of Sr$_2$IrO$_4$ is dependent on the incident angle\cite{Kim2014}. So, we calculate the correlation function for two different incident angles $\theta=8^{\circ}, 85^{\circ}$ using the SU($6$) spin theory, and the results are presented in Fig.~\ref{fig4} b) where the left hand one is for $\theta=8^{\circ}$ and right hand for $\theta=85^{\circ}$. Below $200$~meV, both results exhibit the gapless antiferromagnetic spin waves dispersing up linearly from the $\Gamma$ point, which are consistent with experiments in Sr$_2$IrO$_4$\cite{Gum2009,Kim2012,Kim2014}. Above $200$~meV, a gap of $180$~meV exists arising from the SOC, and the spin-orbital excitations across the gap are ferromagnetic-like spin waves that are parabolic around $\Gamma$ point. Moreover, there is a small gap in the spin-orbital exciton resulting from the splitting in the $t_{2g}$ orbital. We also notice that these spin-orbital exciton modes correspond to a type of SU($N$) bosons in the framework of the SU($N$) spin-wave theory. As for the incident-angle dependence of the spectrum, one can see that the scattering intensity of the low-energy $J_{\rm eff}=1/2$ antiferromagnetic magnon is suppressed heavily, and at the same time the spin-orbital excitations are strongly enhanced for a small incident angle such as $\theta=8^{\circ}$, as shown in the left-hand side in Fig.~\ref{fig4} b). While, an opposite behavior of the spectrum is observed for a large incident angle such as $\theta=85^{\circ}$ (the right-hand side in Fig.~\ref{fig4} b)).
Around the $\Gamma'$ point, the intensity vanishes and only the dispersion of the spectrum is reserved, because the resolution is influenced due to the antiferromagnetic divergence at $\Gamma'$.

The results presented above demonstrate a good performance of the SU($N$) spin wave theory in the study of magnetic orders and dynamics in TMOs. Compared with the SU($2$) spin wave theory, the SU($N$) theory contains more than one type of uncondensed bosons, so that the spin-orbital or multipolar orders and excitations can be captured. Of course, the linear approximation used here involves only single magnon excitations and does not take their interactions into account. So, the broadening and renormalization of the magnonic spectrum are not captured. To study other spin dynamics, such as magnon decay effects\cite{Chernyshev2009,Winter2017}, one should goes beyond the linear order approximation. We note that some modifications of the spin-wave theory \cite{Takahashi1989,Du2015} have been developed in the SU($2$) case, their generalizations to the SU($N$) case deserve further study.

\section{Conclusion}

In summary, we implement the application of the SU($N$) spin wave theory by introducing an efficient local mean field method based on the supercoherent state. The approach is tested firstly by applying to the investigation of magnetic properties in the SU($4$) antiferromagnetic model in a square lattice. We find a long-range Neel order which is consistent with the quantum Monte Carlo simulations, and this order can be interpreted by multipolar orders of $3/2$ spins. We have also calculated the multipolar spin waves of the SU($4$) antiferromagnetic model, to demonstrate the application of SU($N$) spin wave theory in the description of multipolar orderings. Due to the entanglement of spin and orbital degrees of freedom, the multipole-multipole exchange terms are also present in the effective exchange models of spin-orbital Mott insulators. Only if the spin-orbital coupling is large enough that the low-energy physics is confined in Kramers doublet, the effective Hamiltonian will be described by an isospin-$1/2$ model. In this aspect, we examine a toy three-band Hubbard model on a hexagon lattice and find that the Hund's coupling also affects the validity of the isospin-$1/2$ picture when the spin-orbital coupling is below a critical value. Finally, we apply the SU($N$) spin wave theory to two systems of spin-orbital Mott insulators, $\alpha$-RuCl$_3$ and Sr$_2$IrO$_4$. Our results for the magnetic ground states and their low-energy spin dynamics in both systems are consistent with recent experiments. We also obtain the high-energy spin-orbital excitations across the gap in the presence of the spin-orbital coupling.
%Those in Sr$_2$IrO$_4$ are compared to the resonant inelastic X-ray scattering (RIXS) experiments, while those in $\alpha$-RuCl$_3$ wait for a comparison with future experiments.

\begin{acknowledgments}
This work was supported by the National Natural Science Foundation of China (11374138 and 11774152) and National Key Projects for Research and Development of China (Grant No. 2016YFA0300401).
\end{acknowledgments}

\bibliography{SUn}

\begin{thebibliography}{41}%
\makeatletter
\providecommand \@ifxundefined [1]{%
 \@ifx{#1\undefined}
}%
\providecommand \@ifnum [1]{%
 \ifnum #1\expandafter \@firstoftwo
 \else \expandafter \@secondoftwo
 \fi
}%
\providecommand \@ifx [1]{%
 \ifx #1\expandafter \@firstoftwo
 \else \expandafter \@secondoftwo
 \fi
}%
\providecommand \natexlab [1]{#1}%
\providecommand \enquote  [1]{``#1''}%
\providecommand \bibnamefont  [1]{#1}%
\providecommand \bibfnamefont [1]{#1}%
\providecommand \citenamefont [1]{#1}%
\providecommand \href@noop [0]{\@secondoftwo}%
\providecommand \href [0]{\begingroup \@sanitize@url \@href}%
\providecommand \@href[1]{\@@startlink{#1}\@@href}%
\providecommand \@@href[1]{\endgroup#1\@@endlink}%
\providecommand \@sanitize@url [0]{\catcode `\\12\catcode `\$12\catcode
  `\&12\catcode `\#12\catcode `\^12\catcode `\_12\catcode `\%12\relax}%
\providecommand \@@startlink[1]{}%
\providecommand \@@endlink[0]{}%
\providecommand \url  [0]{\begingroup\@sanitize@url \@url }%
\providecommand \@url [1]{\endgroup\@href {#1}{\urlprefix }}%
\providecommand \urlprefix  [0]{URL }%
\providecommand \Eprint [0]{\href }%
\providecommand \doibase [0]{http://dx.doi.org/}%
\providecommand \selectlanguage [0]{\@gobble}%
\providecommand \bibinfo  [0]{\@secondoftwo}%
\providecommand \bibfield  [0]{\@secondoftwo}%
\providecommand \translation [1]{[#1]}%
\providecommand \BibitemOpen [0]{}%
\providecommand \bibitemStop [0]{}%
\providecommand \bibitemNoStop [0]{.\EOS\space}%
\providecommand \EOS [0]{\spacefactor3000\relax}%
\providecommand \BibitemShut  [1]{\csname bibitem#1\endcsname}%
\let\auto@bib@innerbib\@empty
%</preamble>
\bibitem [{\citenamefont {Witczak-Krempa}\ \emph {et~al.}(2014)\citenamefont
  {Witczak-Krempa}, \citenamefont {Chen}, \citenamefont {Kim},\ and\
  \citenamefont {Balents}}]{Witczak-Krempa2014}%
  \BibitemOpen
  \bibfield  {author} {\bibinfo {author} {\bibfnamefont {W.}~\bibnamefont
  {Witczak-Krempa}}, \bibinfo {author} {\bibfnamefont {G.}~\bibnamefont
  {Chen}}, \bibinfo {author} {\bibfnamefont {Y.~B.}\ \bibnamefont {Kim}}, \
  and\ \bibinfo {author} {\bibfnamefont {L.}~\bibnamefont {Balents}},\ }\href
  {\doibase 10.1146/annurev-conmatphys-020911-125138} {\bibfield  {journal}
  {\bibinfo  {journal} {Annu. Rev. Condens. Matter Phys.}\ }\textbf {\bibinfo
  {volume} {5}},\ \bibinfo {pages} {57} (\bibinfo {year} {2014})}\BibitemShut
  {NoStop}%
\bibitem [{\citenamefont {Kim}\ \emph {et~al.}(2008)\citenamefont {Kim},
  \citenamefont {Jin}, \citenamefont {Moon}, \citenamefont {Kim}, \citenamefont
  {Park}, \citenamefont {Leem}, \citenamefont {Yu}, \citenamefont {Noh},
  \citenamefont {Kim}, \citenamefont {Oh}, \citenamefont {Park}, \citenamefont
  {Durairaj}, \citenamefont {Cao},\ and\ \citenamefont {Rotenberg}}]{Kim2008}%
  \BibitemOpen
  \bibfield  {author} {\bibinfo {author} {\bibfnamefont {B.~J.}\ \bibnamefont
  {Kim}}, \bibinfo {author} {\bibfnamefont {H.}~\bibnamefont {Jin}}, \bibinfo
  {author} {\bibfnamefont {S.~J.}\ \bibnamefont {Moon}}, \bibinfo {author}
  {\bibfnamefont {J.-Y.}\ \bibnamefont {Kim}}, \bibinfo {author} {\bibfnamefont
  {B.-G.}\ \bibnamefont {Park}}, \bibinfo {author} {\bibfnamefont {C.~S.}\
  \bibnamefont {Leem}}, \bibinfo {author} {\bibfnamefont {J.}~\bibnamefont
  {Yu}}, \bibinfo {author} {\bibfnamefont {T.~W.}\ \bibnamefont {Noh}},
  \bibinfo {author} {\bibfnamefont {C.}~\bibnamefont {Kim}}, \bibinfo {author}
  {\bibfnamefont {S.-J.}\ \bibnamefont {Oh}}, \bibinfo {author} {\bibfnamefont
  {J.-H.}\ \bibnamefont {Park}}, \bibinfo {author} {\bibfnamefont
  {V.}~\bibnamefont {Durairaj}}, \bibinfo {author} {\bibfnamefont
  {G.}~\bibnamefont {Cao}}, \ and\ \bibinfo {author} {\bibfnamefont
  {E.}~\bibnamefont {Rotenberg}},\ }\href {\doibase
  10.1103/PhysRevLett.101.076402} {\bibfield  {journal} {\bibinfo  {journal}
  {Phys. Rev. Lett.}\ }\textbf {\bibinfo {volume} {101}},\ \bibinfo {pages}
  {076402} (\bibinfo {year} {2008})}\BibitemShut {NoStop}%
\bibitem [{\citenamefont {Kim}\ \emph {et~al.}(2014)\citenamefont {Kim},
  \citenamefont {Daghofer}, \citenamefont {Said}, \citenamefont {Gog},
  \citenamefont {van~den Brink}, \citenamefont {Khaliullin},\ and\
  \citenamefont {Kim}}]{Kim2014}%
  \BibitemOpen
  \bibfield  {author} {\bibinfo {author} {\bibfnamefont {J.}~\bibnamefont
  {Kim}}, \bibinfo {author} {\bibfnamefont {M.}~\bibnamefont {Daghofer}},
  \bibinfo {author} {\bibfnamefont {A.~H.}\ \bibnamefont {Said}}, \bibinfo
  {author} {\bibfnamefont {T.}~\bibnamefont {Gog}}, \bibinfo {author}
  {\bibfnamefont {J.}~\bibnamefont {van~den Brink}}, \bibinfo {author}
  {\bibfnamefont {G.}~\bibnamefont {Khaliullin}}, \ and\ \bibinfo {author}
  {\bibfnamefont {B.~J.}\ \bibnamefont {Kim}},\ }\href {\doibase
  10.1038/ncomms5453} {\bibfield  {journal} {\bibinfo  {journal} {Nat.
  Commun.}\ }\textbf {\bibinfo {volume} {5}},\ \bibinfo {pages} {4453}
  (\bibinfo {year} {2014})}\BibitemShut {NoStop}%
\bibitem [{\citenamefont {Jackeli}\ and\ \citenamefont
  {Khaliullin}(2009)}]{Jackeli2009}%
  \BibitemOpen
  \bibfield  {author} {\bibinfo {author} {\bibfnamefont {G.}~\bibnamefont
  {Jackeli}}\ and\ \bibinfo {author} {\bibfnamefont {G.}~\bibnamefont
  {Khaliullin}},\ }\href {\doibase 10.1103/PhysRevLett.102.017205} {\bibfield
  {journal} {\bibinfo  {journal} {Phys. Rev. Lett.}\ }\textbf {\bibinfo
  {volume} {102}},\ \bibinfo {pages} {017205} (\bibinfo {year}
  {2009})}\BibitemShut {NoStop}%
\bibitem [{\citenamefont {Liu}\ \emph {et~al.}(2011)\citenamefont {Liu},
  \citenamefont {Berlijn}, \citenamefont {Yin}, \citenamefont {Ku},
  \citenamefont {Tsvelik}, \citenamefont {Kim}, \citenamefont {Gretarsson},
  \citenamefont {Singh}, \citenamefont {Gegenwart},\ and\ \citenamefont
  {Hill}}]{Liu2011}%
  \BibitemOpen
  \bibfield  {author} {\bibinfo {author} {\bibfnamefont {X.}~\bibnamefont
  {Liu}}, \bibinfo {author} {\bibfnamefont {T.}~\bibnamefont {Berlijn}},
  \bibinfo {author} {\bibfnamefont {W.-G.}\ \bibnamefont {Yin}}, \bibinfo
  {author} {\bibfnamefont {W.}~\bibnamefont {Ku}}, \bibinfo {author}
  {\bibfnamefont {A.}~\bibnamefont {Tsvelik}}, \bibinfo {author} {\bibfnamefont
  {Y.-J.}\ \bibnamefont {Kim}}, \bibinfo {author} {\bibfnamefont
  {H.}~\bibnamefont {Gretarsson}}, \bibinfo {author} {\bibfnamefont
  {Y.}~\bibnamefont {Singh}}, \bibinfo {author} {\bibfnamefont
  {P.}~\bibnamefont {Gegenwart}}, \ and\ \bibinfo {author} {\bibfnamefont
  {J.~P.}\ \bibnamefont {Hill}},\ }\href {\doibase 10.1103/PhysRevB.83.220403}
  {\bibfield  {journal} {\bibinfo  {journal} {Phys. Rev. B}\ }\textbf {\bibinfo
  {volume} {83}},\ \bibinfo {pages} {220403} (\bibinfo {year}
  {2011})}\BibitemShut {NoStop}%
\bibitem [{\citenamefont {Sears}\ \emph {et~al.}(2015)\citenamefont {Sears},
  \citenamefont {Songvilay}, \citenamefont {Plumb}, \citenamefont {Clancy},
  \citenamefont {Qiu}, \citenamefont {Zhao}, \citenamefont {Parshall},\ and\
  \citenamefont {Kim}}]{Sears2015}%
  \BibitemOpen
  \bibfield  {author} {\bibinfo {author} {\bibfnamefont {J.~S.}\ \bibnamefont
  {Sears}}, \bibinfo {author} {\bibfnamefont {M.}~\bibnamefont {Songvilay}},
  \bibinfo {author} {\bibfnamefont {K.~W.}\ \bibnamefont {Plumb}}, \bibinfo
  {author} {\bibfnamefont {J.~P.}\ \bibnamefont {Clancy}}, \bibinfo {author}
  {\bibfnamefont {Y.}~\bibnamefont {Qiu}}, \bibinfo {author} {\bibfnamefont
  {Y.}~\bibnamefont {Zhao}}, \bibinfo {author} {\bibfnamefont {D.}~\bibnamefont
  {Parshall}}, \ and\ \bibinfo {author} {\bibfnamefont {Y.-J.}\ \bibnamefont
  {Kim}},\ }\href {\doibase 10.1103/PhysRevB.91.144420} {\bibfield  {journal}
  {\bibinfo  {journal} {Phys. Rev. B}\ }\textbf {\bibinfo {volume} {91}},\
  \bibinfo {pages} {144420} (\bibinfo {year} {2015})}\BibitemShut {NoStop}%
\bibitem [{\citenamefont {Banerjee}\ \emph {et~al.}()\citenamefont {Banerjee},
  \citenamefont {Yan}, \citenamefont {Knolle}, \citenamefont {Bridges},
  \citenamefont {Stone}, \citenamefont {Lumsden}, \citenamefont {Mandrus},
  \citenamefont {Tennant}, \citenamefont {Moessner},\ and\ \citenamefont
  {Nagler}}]{Banerjee2016}%
  \BibitemOpen
  \bibfield  {author} {\bibinfo {author} {\bibfnamefont {A.}~\bibnamefont
  {Banerjee}}, \bibinfo {author} {\bibfnamefont {J.}~\bibnamefont {Yan}},
  \bibinfo {author} {\bibfnamefont {J.}~\bibnamefont {Knolle}}, \bibinfo
  {author} {\bibfnamefont {C.~A.}\ \bibnamefont {Bridges}}, \bibinfo {author}
  {\bibfnamefont {M.~B.}\ \bibnamefont {Stone}}, \bibinfo {author}
  {\bibfnamefont {M.~D.}\ \bibnamefont {Lumsden}}, \bibinfo {author}
  {\bibfnamefont {D.~G.}\ \bibnamefont {Mandrus}}, \bibinfo {author}
  {\bibfnamefont {D.~A.}\ \bibnamefont {Tennant}}, \bibinfo {author}
  {\bibfnamefont {R.}~\bibnamefont {Moessner}}, \ and\ \bibinfo {author}
  {\bibfnamefont {S.~E.}\ \bibnamefont {Nagler}},\ }\href
  {http://arxiv.org/abs/1609.00103} {\ }\Eprint
  {http://arxiv.org/abs/1609.00103} {arXiv:1609.00103} \BibitemShut {NoStop}%
\bibitem [{\citenamefont {Ran}\ \emph {et~al.}(2017)\citenamefont {Ran},
  \citenamefont {Wang}, \citenamefont {Wang}, \citenamefont {Dong},
  \citenamefont {Ren}, \citenamefont {Bao}, \citenamefont {Li}, \citenamefont
  {Ma}, \citenamefont {Gan}, \citenamefont {Zhang}, \citenamefont {Park},
  \citenamefont {Deng}, \citenamefont {Danilkin}, \citenamefont {Yu},
  \citenamefont {Li},\ and\ \citenamefont {Wen}}]{Ran2017}%
  \BibitemOpen
  \bibfield  {author} {\bibinfo {author} {\bibfnamefont {K.}~\bibnamefont
  {Ran}}, \bibinfo {author} {\bibfnamefont {J.}~\bibnamefont {Wang}}, \bibinfo
  {author} {\bibfnamefont {W.}~\bibnamefont {Wang}}, \bibinfo {author}
  {\bibfnamefont {Z.-Y.}\ \bibnamefont {Dong}}, \bibinfo {author}
  {\bibfnamefont {X.}~\bibnamefont {Ren}}, \bibinfo {author} {\bibfnamefont
  {S.}~\bibnamefont {Bao}}, \bibinfo {author} {\bibfnamefont {S.}~\bibnamefont
  {Li}}, \bibinfo {author} {\bibfnamefont {Z.}~\bibnamefont {Ma}}, \bibinfo
  {author} {\bibfnamefont {Y.}~\bibnamefont {Gan}}, \bibinfo {author}
  {\bibfnamefont {Y.}~\bibnamefont {Zhang}}, \bibinfo {author} {\bibfnamefont
  {J.~T.}\ \bibnamefont {Park}}, \bibinfo {author} {\bibfnamefont
  {G.}~\bibnamefont {Deng}}, \bibinfo {author} {\bibfnamefont {S.}~\bibnamefont
  {Danilkin}}, \bibinfo {author} {\bibfnamefont {S.-L.}\ \bibnamefont {Yu}},
  \bibinfo {author} {\bibfnamefont {J.-X.}\ \bibnamefont {Li}}, \ and\ \bibinfo
  {author} {\bibfnamefont {J.}~\bibnamefont {Wen}},\ }\href {\doibase
  10.1103/PhysRevLett.118.107203} {\bibfield  {journal} {\bibinfo  {journal}
  {Phys. Rev. Lett.}\ }\textbf {\bibinfo {volume} {118}},\ \bibinfo {pages}
  {107203} (\bibinfo {year} {2017})}\BibitemShut {NoStop}%
\bibitem [{\citenamefont {Biffin}\ \emph
  {et~al.}(2014{\natexlab{a}})\citenamefont {Biffin}, \citenamefont {Johnson},
  \citenamefont {Kimchi}, \citenamefont {Morris}, \citenamefont {Bombardi},
  \citenamefont {Analytis}, \citenamefont {Vishwanath},\ and\ \citenamefont
  {Coldea}}]{Biffin2014}%
  \BibitemOpen
  \bibfield  {author} {\bibinfo {author} {\bibfnamefont {A.}~\bibnamefont
  {Biffin}}, \bibinfo {author} {\bibfnamefont {R.~D.}\ \bibnamefont {Johnson}},
  \bibinfo {author} {\bibfnamefont {I.}~\bibnamefont {Kimchi}}, \bibinfo
  {author} {\bibfnamefont {R.}~\bibnamefont {Morris}}, \bibinfo {author}
  {\bibfnamefont {A.}~\bibnamefont {Bombardi}}, \bibinfo {author}
  {\bibfnamefont {J.~G.}\ \bibnamefont {Analytis}}, \bibinfo {author}
  {\bibfnamefont {A.}~\bibnamefont {Vishwanath}}, \ and\ \bibinfo {author}
  {\bibfnamefont {R.}~\bibnamefont {Coldea}},\ }\href {\doibase
  10.1103/PhysRevLett.113.197201} {\bibfield  {journal} {\bibinfo  {journal}
  {Phys. Rev. Lett.}\ }\textbf {\bibinfo {volume} {113}},\ \bibinfo {pages}
  {197201} (\bibinfo {year} {2014}{\natexlab{a}})}\BibitemShut {NoStop}%
\bibitem [{\citenamefont {Williams}\ \emph {et~al.}(2016)\citenamefont
  {Williams}, \citenamefont {Johnson}, \citenamefont {Freund}, \citenamefont
  {Choi}, \citenamefont {Jesche}, \citenamefont {Kimchi}, \citenamefont
  {Manni}, \citenamefont {Bombardi}, \citenamefont {Manuel}, \citenamefont
  {Gegenwart},\ and\ \citenamefont {Coldea}}]{Williams2016}%
  \BibitemOpen
  \bibfield  {author} {\bibinfo {author} {\bibfnamefont {S.~C.}\ \bibnamefont
  {Williams}}, \bibinfo {author} {\bibfnamefont {R.~D.}\ \bibnamefont
  {Johnson}}, \bibinfo {author} {\bibfnamefont {F.}~\bibnamefont {Freund}},
  \bibinfo {author} {\bibfnamefont {S.}~\bibnamefont {Choi}}, \bibinfo {author}
  {\bibfnamefont {A.}~\bibnamefont {Jesche}}, \bibinfo {author} {\bibfnamefont
  {I.}~\bibnamefont {Kimchi}}, \bibinfo {author} {\bibfnamefont
  {S.}~\bibnamefont {Manni}}, \bibinfo {author} {\bibfnamefont
  {A.}~\bibnamefont {Bombardi}}, \bibinfo {author} {\bibfnamefont
  {P.}~\bibnamefont {Manuel}}, \bibinfo {author} {\bibfnamefont
  {P.}~\bibnamefont {Gegenwart}}, \ and\ \bibinfo {author} {\bibfnamefont
  {R.}~\bibnamefont {Coldea}},\ }\href {\doibase 10.1103/PhysRevB.93.195158}
  {\bibfield  {journal} {\bibinfo  {journal} {Phys. Rev. B}\ }\textbf {\bibinfo
  {volume} {93}},\ \bibinfo {pages} {195158} (\bibinfo {year}
  {2016})}\BibitemShut {NoStop}%
\bibitem [{\citenamefont {Biffin}\ \emph
  {et~al.}(2014{\natexlab{b}})\citenamefont {Biffin}, \citenamefont {Johnson},
  \citenamefont {Choi}, \citenamefont {Freund}, \citenamefont {Manni},
  \citenamefont {Bombardi}, \citenamefont {Manuel}, \citenamefont {Gegenwart},\
  and\ \citenamefont {Coldea}}]{Biffin2014a}%
  \BibitemOpen
  \bibfield  {author} {\bibinfo {author} {\bibfnamefont {A.}~\bibnamefont
  {Biffin}}, \bibinfo {author} {\bibfnamefont {R.~D.}\ \bibnamefont {Johnson}},
  \bibinfo {author} {\bibfnamefont {S.}~\bibnamefont {Choi}}, \bibinfo {author}
  {\bibfnamefont {F.}~\bibnamefont {Freund}}, \bibinfo {author} {\bibfnamefont
  {S.}~\bibnamefont {Manni}}, \bibinfo {author} {\bibfnamefont
  {A.}~\bibnamefont {Bombardi}}, \bibinfo {author} {\bibfnamefont
  {P.}~\bibnamefont {Manuel}}, \bibinfo {author} {\bibfnamefont
  {P.}~\bibnamefont {Gegenwart}}, \ and\ \bibinfo {author} {\bibfnamefont
  {R.}~\bibnamefont {Coldea}},\ }\href {\doibase 10.1103/PhysRevB.90.205116}
  {\bibfield  {journal} {\bibinfo  {journal} {Phys. Rev. B}\ }\textbf {\bibinfo
  {volume} {90}},\ \bibinfo {pages} {205116} (\bibinfo {year}
  {2014}{\natexlab{b}})}\BibitemShut {NoStop}%
\bibitem [{\citenamefont {Kim}\ \emph {et~al.}(2009)\citenamefont {Kim},
  \citenamefont {Ohsumi}, \citenamefont {Komesu}, \citenamefont {Sakai},
  \citenamefont {Morita}, \citenamefont {Takagi},\ and\ \citenamefont
  {Arima}}]{Gum2009}%
  \BibitemOpen
  \bibfield  {author} {\bibinfo {author} {\bibfnamefont {B.~J.}\ \bibnamefont
  {Kim}}, \bibinfo {author} {\bibfnamefont {H.}~\bibnamefont {Ohsumi}},
  \bibinfo {author} {\bibfnamefont {T.}~\bibnamefont {Komesu}}, \bibinfo
  {author} {\bibfnamefont {S.}~\bibnamefont {Sakai}}, \bibinfo {author}
  {\bibfnamefont {T.}~\bibnamefont {Morita}}, \bibinfo {author} {\bibfnamefont
  {H.}~\bibnamefont {Takagi}}, \ and\ \bibinfo {author} {\bibfnamefont
  {T.}~\bibnamefont {Arima}},\ }\href {\doibase 10.1126/science.1167106}
  {\bibfield  {journal} {\bibinfo  {journal} {Science}\ }\textbf {\bibinfo
  {volume} {323}},\ \bibinfo {pages} {1329} (\bibinfo {year}
  {2009})}\BibitemShut {NoStop}%
\bibitem [{\citenamefont {Kim}\ \emph {et~al.}(2012)\citenamefont {Kim},
  \citenamefont {Casa}, \citenamefont {Upton}, \citenamefont {Gog},
  \citenamefont {Kim}, \citenamefont {Mitchell}, \citenamefont
  {Van~Veenendaal}, \citenamefont {Daghofer}, \citenamefont {Van Den~Brink},
  \citenamefont {Khaliullin},\ and\ \citenamefont {Kim}}]{Kim2012}%
  \BibitemOpen
  \bibfield  {author} {\bibinfo {author} {\bibfnamefont {J.}~\bibnamefont
  {Kim}}, \bibinfo {author} {\bibfnamefont {D.}~\bibnamefont {Casa}}, \bibinfo
  {author} {\bibfnamefont {M.~H.}\ \bibnamefont {Upton}}, \bibinfo {author}
  {\bibfnamefont {T.}~\bibnamefont {Gog}}, \bibinfo {author} {\bibfnamefont
  {Y.-J.}\ \bibnamefont {Kim}}, \bibinfo {author} {\bibfnamefont {J.~F.}\
  \bibnamefont {Mitchell}}, \bibinfo {author} {\bibfnamefont {M.}~\bibnamefont
  {Van~Veenendaal}}, \bibinfo {author} {\bibfnamefont {M.}~\bibnamefont
  {Daghofer}}, \bibinfo {author} {\bibfnamefont {J.}~\bibnamefont {Van
  Den~Brink}}, \bibinfo {author} {\bibfnamefont {G.}~\bibnamefont
  {Khaliullin}}, \ and\ \bibinfo {author} {\bibfnamefont {B.~J.}\ \bibnamefont
  {Kim}},\ }\href {\doibase 10.1103/PhysRevLett.108.177003} {\bibfield
  {journal} {\bibinfo  {journal} {Phys. Rev. Lett.}\ }\textbf {\bibinfo
  {volume} {108}},\ \bibinfo {pages} {177003} (\bibinfo {year}
  {2012})}\BibitemShut {NoStop}%
\bibitem [{\citenamefont {Haraldsen}\ and\ \citenamefont
  {Fishman}(2009)}]{Haraldsen2009}%
  \BibitemOpen
  \bibfield  {author} {\bibinfo {author} {\bibfnamefont {J.~T.}\ \bibnamefont
  {Haraldsen}}\ and\ \bibinfo {author} {\bibfnamefont {R.~S.}\ \bibnamefont
  {Fishman}},\ }\href {http://stacks.iop.org/0953-8984/21/i=21/a=216001}
  {\bibfield  {journal} {\bibinfo  {journal} {J. Phys.: Condens. Matter}\
  }\textbf {\bibinfo {volume} {21}},\ \bibinfo {pages} {216001} (\bibinfo
  {year} {2009})}\BibitemShut {NoStop}%
\bibitem [{\citenamefont {Kim}\ \emph {et~al.}(2015)\citenamefont {Kim},
  \citenamefont {Shankar~V.}, \citenamefont {Catuneanu},\ and\ \citenamefont
  {Kee}}]{PhysRevB.91.241110}%
  \BibitemOpen
  \bibfield  {author} {\bibinfo {author} {\bibfnamefont {H.-S.}\ \bibnamefont
  {Kim}}, \bibinfo {author} {\bibfnamefont {V.}~\bibnamefont {Shankar~V.}},
  \bibinfo {author} {\bibfnamefont {A.}~\bibnamefont {Catuneanu}}, \ and\
  \bibinfo {author} {\bibfnamefont {H.-Y.}\ \bibnamefont {Kee}},\ }\href
  {\doibase 10.1103/PhysRevB.91.241110} {\bibfield  {journal} {\bibinfo
  {journal} {Phys. Rev. B}\ }\textbf {\bibinfo {volume} {91}},\ \bibinfo
  {pages} {241110} (\bibinfo {year} {2015})}\BibitemShut {NoStop}%
\bibitem [{\citenamefont {Joshi}\ \emph {et~al.}(1999)\citenamefont {Joshi},
  \citenamefont {Ma}, \citenamefont {Mila}, \citenamefont {Shi},\ and\
  \citenamefont {Zhang}}]{PhysRevB.60.6584}%
  \BibitemOpen
  \bibfield  {author} {\bibinfo {author} {\bibfnamefont {A.}~\bibnamefont
  {Joshi}}, \bibinfo {author} {\bibfnamefont {M.}~\bibnamefont {Ma}}, \bibinfo
  {author} {\bibfnamefont {F.}~\bibnamefont {Mila}}, \bibinfo {author}
  {\bibfnamefont {D.~N.}\ \bibnamefont {Shi}}, \ and\ \bibinfo {author}
  {\bibfnamefont {F.~C.}\ \bibnamefont {Zhang}},\ }\href {\doibase
  10.1103/PhysRevB.60.6584} {\bibfield  {journal} {\bibinfo  {journal} {Phys.
  Rev. B}\ }\textbf {\bibinfo {volume} {60}},\ \bibinfo {pages} {6584}
  (\bibinfo {year} {1999})}\BibitemShut {NoStop}%
\bibitem [{\citenamefont {Bauer}\ \emph {et~al.}(2012)\citenamefont {Bauer},
  \citenamefont {Corboz}, \citenamefont {L\"auchli}, \citenamefont {Messio},
  \citenamefont {Penc}, \citenamefont {Troyer},\ and\ \citenamefont
  {Mila}}]{PhysRevB.85.125116}%
  \BibitemOpen
  \bibfield  {author} {\bibinfo {author} {\bibfnamefont {B.}~\bibnamefont
  {Bauer}}, \bibinfo {author} {\bibfnamefont {P.}~\bibnamefont {Corboz}},
  \bibinfo {author} {\bibfnamefont {A.~M.}\ \bibnamefont {L\"auchli}}, \bibinfo
  {author} {\bibfnamefont {L.}~\bibnamefont {Messio}}, \bibinfo {author}
  {\bibfnamefont {K.}~\bibnamefont {Penc}}, \bibinfo {author} {\bibfnamefont
  {M.}~\bibnamefont {Troyer}}, \ and\ \bibinfo {author} {\bibfnamefont
  {F.}~\bibnamefont {Mila}},\ }\href {\doibase 10.1103/PhysRevB.85.125116}
  {\bibfield  {journal} {\bibinfo  {journal} {Phys. Rev. B}\ }\textbf {\bibinfo
  {volume} {85}},\ \bibinfo {pages} {125116} (\bibinfo {year}
  {2012})}\BibitemShut {NoStop}%
\bibitem [{\citenamefont {Penc}\ \emph {et~al.}(2012)\citenamefont {Penc},
  \citenamefont {Romh\'anyi}, \citenamefont {R\~o\ om}, \citenamefont {Nagel},
  \citenamefont {Antal}, \citenamefont {Feh\'er}, \citenamefont {J\'anossy},
  \citenamefont {Engelkamp}, \citenamefont {Murakawa}, \citenamefont {Tokura},
  \citenamefont {Szaller}, \citenamefont {Bord\'acs},\ and\ \citenamefont
  {K\'ezsm\'arki}}]{PhysRevLett.108.257203}%
  \BibitemOpen
  \bibfield  {author} {\bibinfo {author} {\bibfnamefont {K.}~\bibnamefont
  {Penc}}, \bibinfo {author} {\bibfnamefont {J.}~\bibnamefont {Romh\'anyi}},
  \bibinfo {author} {\bibfnamefont {T.}~\bibnamefont {R\~o\ om}}, \bibinfo
  {author} {\bibfnamefont {U.}~\bibnamefont {Nagel}}, \bibinfo {author}
  {\bibfnamefont {A.}~\bibnamefont {Antal}}, \bibinfo {author} {\bibfnamefont
  {T.}~\bibnamefont {Feh\'er}}, \bibinfo {author} {\bibfnamefont
  {A.}~\bibnamefont {J\'anossy}}, \bibinfo {author} {\bibfnamefont
  {H.}~\bibnamefont {Engelkamp}}, \bibinfo {author} {\bibfnamefont
  {H.}~\bibnamefont {Murakawa}}, \bibinfo {author} {\bibfnamefont
  {Y.}~\bibnamefont {Tokura}}, \bibinfo {author} {\bibfnamefont
  {D.}~\bibnamefont {Szaller}}, \bibinfo {author} {\bibfnamefont
  {S.}~\bibnamefont {Bord\'acs}}, \ and\ \bibinfo {author} {\bibfnamefont
  {I.}~\bibnamefont {K\'ezsm\'arki}},\ }\href {\doibase
  10.1103/PhysRevLett.108.257203} {\bibfield  {journal} {\bibinfo  {journal}
  {Phys. Rev. Lett.}\ }\textbf {\bibinfo {volume} {108}},\ \bibinfo {pages}
  {257203} (\bibinfo {year} {2012})}\BibitemShut {NoStop}%
\bibitem [{\citenamefont {Romh\'anyi}\ and\ \citenamefont
  {Penc}(2012)}]{PhysRevB.86.174428}%
  \BibitemOpen
  \bibfield  {author} {\bibinfo {author} {\bibfnamefont {J.}~\bibnamefont
  {Romh\'anyi}}\ and\ \bibinfo {author} {\bibfnamefont {K.}~\bibnamefont
  {Penc}},\ }\href {\doibase 10.1103/PhysRevB.86.174428} {\bibfield  {journal}
  {\bibinfo  {journal} {Phys. Rev. B}\ }\textbf {\bibinfo {volume} {86}},\
  \bibinfo {pages} {174428} (\bibinfo {year} {2012})}\BibitemShut {NoStop}%
\bibitem [{\citenamefont {Muniz}\ \emph {et~al.}(2014)\citenamefont {Muniz},
  \citenamefont {Kato},\ and\ \citenamefont {Batista}}]{Muniz2014}%
  \BibitemOpen
  \bibfield  {author} {\bibinfo {author} {\bibfnamefont {R.~A.}\ \bibnamefont
  {Muniz}}, \bibinfo {author} {\bibfnamefont {Y.}~\bibnamefont {Kato}}, \ and\
  \bibinfo {author} {\bibfnamefont {C.~D.}\ \bibnamefont {Batista}},\ }\href
  {\doibase 10.1093/ptep/ptu109} {\bibfield  {journal} {\bibinfo  {journal}
  {Prog. Theor. Exp. Phys.}\ }\textbf {\bibinfo {volume} {2014}},\ \bibinfo
  {pages} {83I01} (\bibinfo {year} {2014})}\BibitemShut {NoStop}%
\bibitem [{\citenamefont {Fatyga}\ \emph {et~al.}(1991)\citenamefont {Fatyga},
  \citenamefont {Kosteleck\'{y}}, \citenamefont {Nieto},\ and\ \citenamefont
  {Truax}}]{Fatyga1991}%
  \BibitemOpen
  \bibfield  {author} {\bibinfo {author} {\bibfnamefont {B.~W.}\ \bibnamefont
  {Fatyga}}, \bibinfo {author} {\bibfnamefont {V.~A.}\ \bibnamefont
  {Kosteleck\'{y}}}, \bibinfo {author} {\bibfnamefont {M.~M.}\ \bibnamefont
  {Nieto}}, \ and\ \bibinfo {author} {\bibfnamefont {D.~R.}\ \bibnamefont
  {Truax}},\ }\href {\doibase 10.1103/PhysRevD.43.1403} {\bibfield  {journal}
  {\bibinfo  {journal} {Phys. Rev. D}\ }\textbf {\bibinfo {volume} {43}},\
  \bibinfo {pages} {1403} (\bibinfo {year} {1991})}\BibitemShut {NoStop}%
\bibitem [{\citenamefont {Qi}\ and\ \citenamefont {Xu}(2008)}]{Qi2008a}%
  \BibitemOpen
  \bibfield  {author} {\bibinfo {author} {\bibfnamefont {Y.}~\bibnamefont
  {Qi}}\ and\ \bibinfo {author} {\bibfnamefont {C.}~\bibnamefont {Xu}},\ }\href
  {\doibase 10.1103/PhysRevB.78.014410} {\bibfield  {journal} {\bibinfo
  {journal} {Phys. Rev. B}\ }\textbf {\bibinfo {volume} {78}},\ \bibinfo
  {pages} {014410} (\bibinfo {year} {2008})}\BibitemShut {NoStop}%
\bibitem [{\citenamefont {Wu}\ \emph {et~al.}(2003)\citenamefont {Wu},
  \citenamefont {Hu},\ and\ \citenamefont {Zhang}}]{Wu2003}%
  \BibitemOpen
  \bibfield  {author} {\bibinfo {author} {\bibfnamefont {C.}~\bibnamefont
  {Wu}}, \bibinfo {author} {\bibfnamefont {J.-P.}\ \bibnamefont {Hu}}, \ and\
  \bibinfo {author} {\bibfnamefont {S.-C.}\ \bibnamefont {Zhang}},\ }\href
  {\doibase 10.1103/PhysRevLett.91.186402} {\bibfield  {journal} {\bibinfo
  {journal} {Phys. Rev. Lett.}\ }\textbf {\bibinfo {volume} {91}},\ \bibinfo
  {pages} {186402} (\bibinfo {year} {2003})}\BibitemShut {NoStop}%
\bibitem [{\citenamefont {Hung}\ \emph {et~al.}(2011)\citenamefont {Hung},
  \citenamefont {Wang},\ and\ \citenamefont {Wu}}]{Hung2011}%
  \BibitemOpen
  \bibfield  {author} {\bibinfo {author} {\bibfnamefont {H.-H.}\ \bibnamefont
  {Hung}}, \bibinfo {author} {\bibfnamefont {Y.}~\bibnamefont {Wang}}, \ and\
  \bibinfo {author} {\bibfnamefont {C.}~\bibnamefont {Wu}},\ }\href {\doibase
  10.1103/PhysRevB.84.054406} {\bibfield  {journal} {\bibinfo  {journal} {Phys.
  Rev. B}\ }\textbf {\bibinfo {volume} {84}},\ \bibinfo {pages} {054406}
  (\bibinfo {year} {2011})}\BibitemShut {NoStop}%
\bibitem [{\citenamefont {Luo}\ \emph {et~al.}(1993)\citenamefont {Luo},
  \citenamefont {Trammell},\ and\ \citenamefont {Hannon}}]{Luo1993}%
  \BibitemOpen
  \bibfield  {author} {\bibinfo {author} {\bibfnamefont {J.}~\bibnamefont
  {Luo}}, \bibinfo {author} {\bibfnamefont {G.}~\bibnamefont {Trammell}}, \
  and\ \bibinfo {author} {\bibfnamefont {J.}~\bibnamefont {Hannon}},\ }\href
  {\doibase 10.1103/PhysRevLett.71.287} {\bibfield  {journal} {\bibinfo
  {journal} {Phys. Rev. Lett.}\ }\textbf {\bibinfo {volume} {71}},\ \bibinfo
  {pages} {287} (\bibinfo {year} {1993})}\BibitemShut {NoStop}%
\bibitem [{\citenamefont {Arovas}\ and\ \citenamefont
  {Auerbach}(1988)}]{Arovas1988}%
  \BibitemOpen
  \bibfield  {author} {\bibinfo {author} {\bibfnamefont {D.~P.}\ \bibnamefont
  {Arovas}}\ and\ \bibinfo {author} {\bibfnamefont {A.}~\bibnamefont
  {Auerbach}},\ }\href {\doibase 10.1103/PhysRevB.38.316} {\bibfield  {journal}
  {\bibinfo  {journal} {Phys. Rev. B}\ }\textbf {\bibinfo {volume} {38}},\
  \bibinfo {pages} {316} (\bibinfo {year} {1988})}\BibitemShut {NoStop}%
\bibitem [{\citenamefont {White}\ \emph {et~al.}(1965)\citenamefont {White},
  \citenamefont {Sparks},\ and\ \citenamefont {Ortenburger}}]{White1965}%
  \BibitemOpen
  \bibfield  {author} {\bibinfo {author} {\bibfnamefont {R.~M.}\ \bibnamefont
  {White}}, \bibinfo {author} {\bibfnamefont {M.}~\bibnamefont {Sparks}}, \
  and\ \bibinfo {author} {\bibfnamefont {I.}~\bibnamefont {Ortenburger}},\
  }\href {\doibase 10.1103/PhysRev.139.A450} {\bibfield  {journal} {\bibinfo
  {journal} {Phys. Rev.}\ }\textbf {\bibinfo {volume} {139}},\ \bibinfo {pages}
  {A450} (\bibinfo {year} {1965})}\BibitemShut {NoStop}%
\bibitem [{\citenamefont {Colpa}(1978)}]{Colpa1978}%
  \BibitemOpen
  \bibfield  {author} {\bibinfo {author} {\bibfnamefont {J.}~\bibnamefont
  {Colpa}},\ }\href {\doibase 10.1016/0378-4371(78)90160-7} {\bibfield
  {journal} {\bibinfo  {journal} {Physica A}\ }\textbf {\bibinfo {volume}
  {93}},\ \bibinfo {pages} {327} (\bibinfo {year} {1978})}\BibitemShut
  {NoStop}%
\bibitem [{\citenamefont {Harada}\ \emph {et~al.}(2003)\citenamefont {Harada},
  \citenamefont {Kawashima},\ and\ \citenamefont {Troyer}}]{Harada2003}%
  \BibitemOpen
  \bibfield  {author} {\bibinfo {author} {\bibfnamefont {K.}~\bibnamefont
  {Harada}}, \bibinfo {author} {\bibfnamefont {N.}~\bibnamefont {Kawashima}}, \
  and\ \bibinfo {author} {\bibfnamefont {M.}~\bibnamefont {Troyer}},\ }\href
  {\doibase 10.1103/PhysRevLett.90.117203} {\bibfield  {journal} {\bibinfo
  {journal} {Phys. Rev. Lett.}\ }\textbf {\bibinfo {volume} {90}},\ \bibinfo
  {pages} {117203} (\bibinfo {year} {2003})}\BibitemShut {NoStop}%
\bibitem [{\citenamefont {Nataf}\ \emph {et~al.}(2016)\citenamefont {Nataf},
  \citenamefont {Lajk\'{o}}, \citenamefont {Corboz}, \citenamefont
  {L\"{a}uchli}, \citenamefont {Penc},\ and\ \citenamefont {Mila}}]{Nataf2016}%
  \BibitemOpen
  \bibfield  {author} {\bibinfo {author} {\bibfnamefont {P.}~\bibnamefont
  {Nataf}}, \bibinfo {author} {\bibfnamefont {M.}~\bibnamefont {Lajk\'{o}}},
  \bibinfo {author} {\bibfnamefont {P.}~\bibnamefont {Corboz}}, \bibinfo
  {author} {\bibfnamefont {A.~M.}\ \bibnamefont {L\"{a}uchli}}, \bibinfo
  {author} {\bibfnamefont {K.}~\bibnamefont {Penc}}, \ and\ \bibinfo {author}
  {\bibfnamefont {F.}~\bibnamefont {Mila}},\ }\href {\doibase
  10.1103/PhysRevB.93.201113} {\bibfield  {journal} {\bibinfo  {journal} {Phys.
  Rev. B}\ }\textbf {\bibinfo {volume} {93}},\ \bibinfo {pages} {201113}
  (\bibinfo {year} {2016})}\BibitemShut {NoStop}%
\bibitem [{\citenamefont {Zhao}\ \emph {et~al.}(2012)\citenamefont {Zhao},
  \citenamefont {Xu}, \citenamefont {Chen}, \citenamefont {Wei}, \citenamefont
  {Qin}, \citenamefont {Zhang},\ and\ \citenamefont {Xiang}}]{Zhao2012}%
  \BibitemOpen
  \bibfield  {author} {\bibinfo {author} {\bibfnamefont {H.~H.}\ \bibnamefont
  {Zhao}}, \bibinfo {author} {\bibfnamefont {C.}~\bibnamefont {Xu}}, \bibinfo
  {author} {\bibfnamefont {Q.~N.}\ \bibnamefont {Chen}}, \bibinfo {author}
  {\bibfnamefont {Z.~C.}\ \bibnamefont {Wei}}, \bibinfo {author} {\bibfnamefont
  {M.~P.}\ \bibnamefont {Qin}}, \bibinfo {author} {\bibfnamefont {G.~M.}\
  \bibnamefont {Zhang}}, \ and\ \bibinfo {author} {\bibfnamefont
  {T.}~\bibnamefont {Xiang}},\ }\href {\doibase 10.1103/PhysRevB.85.134416}
  {\bibfield  {journal} {\bibinfo  {journal} {Phys. Rev. B}\ }\textbf {\bibinfo
  {volume} {85}},\ \bibinfo {pages} {134416} (\bibinfo {year}
  {2012})}\BibitemShut {NoStop}%
\bibitem [{\citenamefont {Wang}\ \emph {et~al.}(2017)\citenamefont {Wang},
  \citenamefont {Dong}, \citenamefont {Yu},\ and\ \citenamefont
  {Li}}]{Wang2017}%
  \BibitemOpen
  \bibfield  {author} {\bibinfo {author} {\bibfnamefont {W.}~\bibnamefont
  {Wang}}, \bibinfo {author} {\bibfnamefont {Z.-Y.}\ \bibnamefont {Dong}},
  \bibinfo {author} {\bibfnamefont {S.-L.}\ \bibnamefont {Yu}}, \ and\ \bibinfo
  {author} {\bibfnamefont {J.-X.}\ \bibnamefont {Li}},\ }\href {\doibase
  10.1103/PhysRevB.96.115103} {\bibfield  {journal} {\bibinfo  {journal} {Phys.
  Rev. B}\ }\textbf {\bibinfo {volume} {96}},\ \bibinfo {pages} {115103}
  (\bibinfo {year} {2017})}\BibitemShut {NoStop}%
\bibitem [{\citenamefont {Koitzsch}\ \emph {et~al.}(2016)\citenamefont
  {Koitzsch}, \citenamefont {Habenicht}, \citenamefont {M\"uller},
  \citenamefont {Knupfer}, \citenamefont {B\"uchner}, \citenamefont {Kandpal},
  \citenamefont {van~den Brink}, \citenamefont {Nowak}, \citenamefont
  {Isaeva},\ and\ \citenamefont {Doert}}]{PhysRevLett.117.126403}%
  \BibitemOpen
  \bibfield  {author} {\bibinfo {author} {\bibfnamefont {A.}~\bibnamefont
  {Koitzsch}}, \bibinfo {author} {\bibfnamefont {C.}~\bibnamefont {Habenicht}},
  \bibinfo {author} {\bibfnamefont {E.}~\bibnamefont {M\"uller}}, \bibinfo
  {author} {\bibfnamefont {M.}~\bibnamefont {Knupfer}}, \bibinfo {author}
  {\bibfnamefont {B.}~\bibnamefont {B\"uchner}}, \bibinfo {author}
  {\bibfnamefont {H.~C.}\ \bibnamefont {Kandpal}}, \bibinfo {author}
  {\bibfnamefont {J.}~\bibnamefont {van~den Brink}}, \bibinfo {author}
  {\bibfnamefont {D.}~\bibnamefont {Nowak}}, \bibinfo {author} {\bibfnamefont
  {A.}~\bibnamefont {Isaeva}}, \ and\ \bibinfo {author} {\bibfnamefont
  {T.}~\bibnamefont {Doert}},\ }\href {\doibase 10.1103/PhysRevLett.117.126403}
  {\bibfield  {journal} {\bibinfo  {journal} {Phys. Rev. Lett.}\ }\textbf
  {\bibinfo {volume} {117}},\ \bibinfo {pages} {126403} (\bibinfo {year}
  {2016})}\BibitemShut {NoStop}%
\bibitem [{\citenamefont {Sandilands}\ \emph {et~al.}(2016)\citenamefont
  {Sandilands}, \citenamefont {Tian}, \citenamefont {Reijnders}, \citenamefont
  {Kim}, \citenamefont {Plumb}, \citenamefont {Kim}, \citenamefont {Kee},\ and\
  \citenamefont {Burch}}]{PhysRevB.93.075144}%
  \BibitemOpen
  \bibfield  {author} {\bibinfo {author} {\bibfnamefont {L.~J.}\ \bibnamefont
  {Sandilands}}, \bibinfo {author} {\bibfnamefont {Y.}~\bibnamefont {Tian}},
  \bibinfo {author} {\bibfnamefont {A.~A.}\ \bibnamefont {Reijnders}}, \bibinfo
  {author} {\bibfnamefont {H.-S.}\ \bibnamefont {Kim}}, \bibinfo {author}
  {\bibfnamefont {K.~W.}\ \bibnamefont {Plumb}}, \bibinfo {author}
  {\bibfnamefont {Y.-J.}\ \bibnamefont {Kim}}, \bibinfo {author} {\bibfnamefont
  {H.-Y.}\ \bibnamefont {Kee}}, \ and\ \bibinfo {author} {\bibfnamefont
  {K.~S.}\ \bibnamefont {Burch}},\ }\href {\doibase 10.1103/PhysRevB.93.075144}
  {\bibfield  {journal} {\bibinfo  {journal} {Phys. Rev. B}\ }\textbf {\bibinfo
  {volume} {93}},\ \bibinfo {pages} {075144} (\bibinfo {year}
  {2016})}\BibitemShut {NoStop}%
\bibitem [{\citenamefont {Winter}\ \emph {et~al.}(2016)\citenamefont {Winter},
  \citenamefont {Li}, \citenamefont {Jeschke},\ and\ \citenamefont
  {Valent\'{\i}}}]{PhysRevB.93.214431}%
  \BibitemOpen
  \bibfield  {author} {\bibinfo {author} {\bibfnamefont {S.~M.}\ \bibnamefont
  {Winter}}, \bibinfo {author} {\bibfnamefont {Y.}~\bibnamefont {Li}}, \bibinfo
  {author} {\bibfnamefont {H.~O.}\ \bibnamefont {Jeschke}}, \ and\ \bibinfo
  {author} {\bibfnamefont {R.}~\bibnamefont {Valent\'{\i}}},\ }\href {\doibase
  10.1103/PhysRevB.93.214431} {\bibfield  {journal} {\bibinfo  {journal} {Phys.
  Rev. B}\ }\textbf {\bibinfo {volume} {93}},\ \bibinfo {pages} {214431}
  (\bibinfo {year} {2016})}\BibitemShut {NoStop}%
\bibitem [{\citenamefont {Watanabe}\ \emph {et~al.}(2010)\citenamefont
  {Watanabe}, \citenamefont {Shirakawa},\ and\ \citenamefont
  {Yunoki}}]{Watanabe2010}%
  \BibitemOpen
  \bibfield  {author} {\bibinfo {author} {\bibfnamefont {H.}~\bibnamefont
  {Watanabe}}, \bibinfo {author} {\bibfnamefont {T.}~\bibnamefont {Shirakawa}},
  \ and\ \bibinfo {author} {\bibfnamefont {S.}~\bibnamefont {Yunoki}},\ }\href
  {\doibase 10.1103/PhysRevLett.105.216410} {\bibfield  {journal} {\bibinfo
  {journal} {Phys. Rev. Lett.}\ }\textbf {\bibinfo {volume} {105}},\ \bibinfo
  {pages} {216410} (\bibinfo {year} {2010})}\BibitemShut {NoStop}%
\bibitem [{\citenamefont {Wang}\ \emph {et~al.}(2015)\citenamefont {Wang},
  \citenamefont {Yu},\ and\ \citenamefont {Li}}]{Wang2015}%
  \BibitemOpen
  \bibfield  {author} {\bibinfo {author} {\bibfnamefont {H.}~\bibnamefont
  {Wang}}, \bibinfo {author} {\bibfnamefont {S.-L.}\ \bibnamefont {Yu}}, \ and\
  \bibinfo {author} {\bibfnamefont {J.-X.}\ \bibnamefont {Li}},\ }\href
  {\doibase 10.1103/PhysRevB.91.165138} {\bibfield  {journal} {\bibinfo
  {journal} {Phys. Rev. B}\ }\textbf {\bibinfo {volume} {91}},\ \bibinfo
  {pages} {165138} (\bibinfo {year} {2015})}\BibitemShut {NoStop}%
\bibitem [{\citenamefont {Chernyshev}\ and\ \citenamefont
  {Zhitomirsky}(2009)}]{Chernyshev2009}%
  \BibitemOpen
  \bibfield  {author} {\bibinfo {author} {\bibfnamefont {A.~L.}\ \bibnamefont
  {Chernyshev}}\ and\ \bibinfo {author} {\bibfnamefont {M.~E.}\ \bibnamefont
  {Zhitomirsky}},\ }\href {\doibase 10.1103/PhysRevB.79.144416} {\bibfield
  {journal} {\bibinfo  {journal} {Phys. Rev. B}\ }\textbf {\bibinfo {volume}
  {79}},\ \bibinfo {pages} {144416} (\bibinfo {year} {2009})}\BibitemShut
  {NoStop}%
\bibitem [{\citenamefont {Winter}\ \emph {et~al.}()\citenamefont {Winter},
  \citenamefont {Riedl}, \citenamefont {Honecker},\ and\ \citenamefont
  {Valenti}}]{Winter2017}%
  \BibitemOpen
  \bibfield  {author} {\bibinfo {author} {\bibfnamefont {S.~M.}\ \bibnamefont
  {Winter}}, \bibinfo {author} {\bibfnamefont {K.}~\bibnamefont {Riedl}},
  \bibinfo {author} {\bibfnamefont {A.}~\bibnamefont {Honecker}}, \ and\
  \bibinfo {author} {\bibfnamefont {R.}~\bibnamefont {Valenti}},\ }\href
  {http://arxiv.org/abs/1702.08466} {\ }\Eprint
  {http://arxiv.org/abs/1702.08466} {arXiv:1702.08466} \BibitemShut {NoStop}%
\bibitem [{\citenamefont {Takahashi}(1989)}]{Takahashi1989}%
  \BibitemOpen
  \bibfield  {author} {\bibinfo {author} {\bibfnamefont {M.}~\bibnamefont
  {Takahashi}},\ }\href {\doibase 10.1103/PhysRevB.40.2494} {\bibfield
  {journal} {\bibinfo  {journal} {Phys. Rev. B}\ }\textbf {\bibinfo {volume}
  {40}},\ \bibinfo {pages} {2494} (\bibinfo {year} {1989})}\BibitemShut
  {NoStop}%
\bibitem [{\citenamefont {Du}\ \emph {et~al.}(2015)\citenamefont {Du},
  \citenamefont {Liu}, \citenamefont {Xie}, \citenamefont {Wang},\ and\
  \citenamefont {Liu}}]{Du2015}%
  \BibitemOpen
  \bibfield  {author} {\bibinfo {author} {\bibfnamefont {Z.~Z.}\ \bibnamefont
  {Du}}, \bibinfo {author} {\bibfnamefont {H.~M.}\ \bibnamefont {Liu}},
  \bibinfo {author} {\bibfnamefont {Y.~L.}\ \bibnamefont {Xie}}, \bibinfo
  {author} {\bibfnamefont {Q.~H.}\ \bibnamefont {Wang}}, \ and\ \bibinfo
  {author} {\bibfnamefont {J.-M.}\ \bibnamefont {Liu}},\ }\href {\doibase
  10.1103/PhysRevB.92.214409} {\bibfield  {journal} {\bibinfo  {journal} {Phys.
  Rev. B}\ }\textbf {\bibinfo {volume} {92}},\ \bibinfo {pages} {214409}
  (\bibinfo {year} {2015})}\BibitemShut {NoStop}%
\end{thebibliography}%

\end{document}